# Highly-stable, eco-friendly and selective $Cs_2AgBiBr_6$ perovskite-based ozone sensor

*Aikaterini Argyrou[#,$], Rafaela Maria Giappa[^], Emmanouil Gagaoudakis[#], Vassilios Binas[#,&], Ioannis Remediakis[#,^], Konstantinos Brintakis[#]*, Athanasia Kostopoulou[#]*, Emmanuel Stratakis[#]**

[#]Institute of Electronic Structure and Laser
Foundation for Research and Technology-Hellas
Vassilika Vouton, 70013, Heraklion, Greece
E-mail: kbrin@iesl.forth.gr; akosto@iesl.forth.gr; stratak@iesl.forth.gr

[$]Department of Chemistry
University of Crete
Vassilika Vouton, 70013, Heraklion, Greece

[^]Department of Material Science and Technology
University of Crete
Vassilika Vouton, 70013, Heraklion, Greece

[&]Department of Chemistry
Aristotle University of Thessaloniki
Thessaloniki, 54633, Greece

## Abstract

Lead halide perovskites have attracted considerable attention as potential gas sensing elements due to their unique ability to detect and respond to an external stimulus with measurable electrical or optical signals. The distinctive characteristics of these materials lie in their ability to operate upon gas exposure at room temperature. However, the presence of lead (Pb) poses a serious challenge for their widespread application and commercialization, owing to its toxicity. To address this issue, lead-free $Cs_2AgBiBr_6$ double perovskite is explored as a promising alternative sensing element for room temperature gas detection. This sensor can be regarded as eco-friendly for the following reasons: it does not contain Pb; it is synthesized at room temperature without using strong organic solvents; and it operates and recovers at room temperature without heating or UV irradiation and with a very low input voltage (0.1V), thus reducing overall energy consumption. The influence of the perovskite morphology on the ozone ($O_3$) sensing performance is investigated, as well as the sensing stability under varying conditions of time, humidity and temperature. Notably, the sensors' exceptional selectivity for $O_3$ over other gases and the interaction between the gases and the perovskite surface is confirmed through experimental and first-principles calculations.

# 1. Introduction

The problem of indoor and outdoor air quality has become more critical in a world characterized by rapid industrialization and urbanization.[1,2] The increase in air pollution levels, caused by emissions of toxic gases and volatile organic compounds (VOCs), represents a significant risk to public health and environment.[3] Consequently, various gas sensors have been employed for the identification of these hazardous substances, facilitating the real-time monitoring and protection of both human health and environmental well-being.[4,5] The versatility of gas sensing technologies stems from the development of materials that selectively interact with specific target gases. Among these sensing elements, metal oxide semiconductors are the most prevalent due to their fast response and remarkable sensitivity to harmful gases.[6] However, such materials often require lengthy synthesis processes and high operating temperatures, which can potentially compromise their long-term stability and increase energy consumption.[7]

The emergence of Pb-halide perovskites is considered noteworthy in the field of efficient gas sensing elements. These materials are well-known for their unique optoelectronic properties, enabling applications such as LEDs and photovoltaics.[8,9] Moreover, they possess features such as sensitive surfaces, abundant active sites, and controllable defect density, which facilitate the selective detection of gaseous compounds at room temperature.[10,11,12] Several Pb-halide perovskite materials, including both organic-inorganic and all-inorganic, have been employed for the detection of various gases such as molecular oxygen ($O_2$), nitrogen dioxide ($NO_2$) and ammonia ($NH_3$), exhibiting exceptional sensitivity and response.[13,14,15] Nevertheless, despite their outstanding sensing performance, concerns regarding the toxicity of Pb hinder their commercial exploitation, as even low concentrations of Pb can have severe implications for both human health and environment. [16,17] Pb exposure is associated with various health issues, including neurological and cardiovascular problems in humans, while environmental Pb exposure can result in soil and water contamination, harming ecosystem and wildlife.[18,19]

In response to these concerns, the scientific community has shifted its focus to designing environmentally friendly, Pb-free halide perovskites. Pb-free perovskites are obtained by the substitution of $Pb^{2+}$ ions with alternative non-toxic divalent metals such as $Sn^{2+}$, or with isoelectronic trivalent elements, like $Bi^{3+}$ for enhanced stability.[20] Meanwhile, the well-known double perovskites are derived from the incorporation of a monovalent and a trivalent metallic element.[21] An obvious choice towards this end would be to combine trivalent Bi with a monovalent noble metal, such as Ag. Although research on the synthesis and optical properties of these materials has expanded, reports on the detection of gaseous compounds using such materials are very limited, focusing only on the detection of $NO_2$, $NH_3$ and NO.[22,23,24,25,26]

Despite the great potential of Pb-free metal halide perovskites in detecting both reducing and oxidizing gases, to the best of our knowledge, there have been no reports of their capability to detect ozone ($O_3$) gas. $O_3$ is a highly oxidizing agent and is extensively used in many applications, including water purification, food processing and medical sterilization.[27,28,29] On the other hand, despite its numerous beneficial uses, $O_3$ is considered a greenhouse gas and, when its level exceeds a certain threshold value, it becomes harmful to human health. Long-term exposure to $O_3$ is related to headaches, eye irritation, respiratory infections, and lung damage.[30] Therefore, there is an urgent need for inexpensive $O_3$ sensors with exceptional sensitivity to support air-quality monitoring and at the same time be compatible with Internet of Things (IoT) devices.

The present study contributes to the gas sensing technology by introducing environmentally friendly perovskites that are efficient, stable under harsh conditions, and cost-effective due to their quick room-temperature synthesis. Specifically, ligand-free $Cs_2AgBiBr_6$ double perovskites in the form of microsheets, faceted microflowers, and spherical microflowers were synthesized using a precipitation method under ambient conditions. This study further elaborates on the optimal morphology and size for accurate detection and selectivity in various gases, combining experimental data with atomistic simulations to identify specific adsorption sites and surface interactions. The sensing performance, particularly under $O_3$ exposure, and the stability at high temperatures and relative humidity levels are demonstrated. The microsheet-based perovskite sensor was distinguished for its full reversibility and selectivity towards $O_3$ among the NO, $H_2$, $CO_2$, and $CH_4$ gases. This selectivity is further validated by atomistic simulations, which confirm the experimental observations and provide insights into the active sites on the perovskite surface and the underlying sensing mechanisms of $Cs_2AgBiBr_6$ double perovskites against all the tested gases.

# 2. Experimental section

## 2.1 Materials

Chemicals: The chemical precursors CsBr (99.999% trace metal basis), AgBr (99.9%- Ag), and $BiBr_3$ (99.999%, anhydrous, metals basis) were purchased from Sigma Aldrich, Strem Chemicals and Thermo Fisher Scientific, respectively, and used without further treatment before the synthesis of metal halide perovskites. The solvents, Dimethyl Sulfoxide (DMSO, ≥99.9%, anhydrous) and ethanol (≥99.8%, absolute) were purchased from Sigma- Aldrich and Fisher Scientific, respectively.

## 2.2 Syntheses of $Cs_2AgBiBr_6$ microflowers and microsheets

The precursor solution was prepared by dissolving 0.5 mmol of CsBr powder, 0.25 mmol of AgBr powder, and 0.25 mmol of $BiBr_3$ in 10 ml of DMSO under ambient conditions. The ligand-free $Cs_2AgBiBr_6$ microsheets were synthesized through a room-temperature precipitation approach. Specifically, 1 ml of the precursor solution was introduced into 10 ml of ethanol, resulting in the mixture’s color to instantly turn yellow, indicative of rapid crystal nucleation and growth. After a 20-minute precipitation period, the precipitate was collected, deposited onto electrodes for sensing measurements, and subsequently dried under vacuum. Concurrently, the supernatant contained spherical microflowers, which were also deposited onto electrodes and dried under vacuum.

In contrast, the $Cs_2AgBiBr_6$ faceted microflowers were grown directly on electrodes upon heating. A few μl of the precursor solution were deposited on preheated electrodes at 80°C (heating plate temperature set point) and maintained at this temperature for 20 minutes to facilitate growth.

### 2.3 Morphological, structural and chemical characterization of microcrystals

Field-Emission Scanning Electron Microscopy (FESEM, JEOL JSM IT700HR) working at 20 kV was employed to analyze the surface morphology of the perovskites. Energy Dispersive X-Ray Spectroscopy (EDS, Dry SD60 detector) was utilized for the elemental analysis of the prepared samples. The crystal structures were investigated by X-Ray Diffraction (XRD, Bruker AXS D8 Advance copper anode diffractometer) over the 2θ collection range of 10° to 50° with scan rate of 0.02°/s, using a monochromatic Cu Kα radiation source (λ=1.54056 Å). Temperature-dependent XRD measurements were performed over the same 2θ range to study the structural evolution of the material under varying thermal conditions. The samples were subjected to controlled heating at room-temperature, 100° and 200 °C. The absorbance spectra of all fabricated materials were measured by UV-Visible spectroscopy (Perkin Elmer Lambda 950 UV/Vis/NIR) in the range of 380 nm to 650 nm. The chemical compositions and electronic states of perovskites were determined by X-Ray Photoelectron Spectroscopy (XPS, SPECS-Germany, FlexMod) equipped with a PHOIBOS 100 1D-DLD energy analyzer and an Al Kα monochromatic X-Ray source (1486.7 eV) operated with 200 W and 15 kV. The spectra that were recorded were the following: survey spectra, C 1s, Cs 3d, Ag 3d, Bi 3d, Br 4d and O 1s. The survey and the high-resolution spectra were recorded at a pass energy of 30 eV. All the binding energies in XPS spectra were calibrated using the C 1s at 284.8 eV as a reference. The data analysis was performed with SpecsLab Prodigy and CasaXPS (Casa software Ltd, DEMO version). A Tougaard baseline was used in combination with a Gaussian-Lorentzian function for the high-resolution spectra deconvolution, which was executed by OriginPro 2016 software.

### 2.4 Gas-sensing measurements and analysis

The gas-sensing measurements were conducted at room temperature, in a custom-made gas sensing chamber, providing a controlled environment for the electrical assessment of the sensors. The gas sensing setup was consisted of gas suppliers connected with mass flow controllers, coupled with a stainless-steel chamber which was initially evacuated down to $10^{-3}$ mbar. To evaluate the $O_3$ sensing capability of all fabricated materials, a commercial $O_3$ analyzer (Thermo Electron Corporation, Model 49i) was employed. The analyzer was used to supply and record accurately controlled $O_3$ concentrations that were introduced in the chamber at a flow rate of 500 sccm (standard cm$^3$/min) flow. Electrical current measurements over time were executed using an electrometer (Keithley 6517A) by applying a constant voltage. The values of the voltage ranged from 0.1 V for the microsheets and faceted micro-flowers to 1 V for the spherical microflowers. The sensing process was initiated by exposing the sensors to $O_3$ gas for 150 s, followed by a 200 s recovery with synthetic air. All samples were exposed to different $O_3$ concentrations ranging from 2300 ppb down to 4 ppb. During the experimental procedure, the pressure in the chamber maintained constant to 600 mbar.

The impact of temperature, combined with $O_3$ gas, was investigated by replacing the custom chamber with a commercially available Ceramic Heater Type Micro Probe System (MPS-CH, MVPS III-PTTC, NEXTRON), capable to operate within the range of room temperature up to 450 °C. The influence of relative humidity (RH) was also studied by connecting the MPS-CH with a humidity controller. Humidity was provided by injecting deionized (DI) water in the humidifier and the levels of 30, 45, 50 and 70% of RH were studied.

The performance of the sensors for each gas is assessed through their response (R), defined as:

$$R = \frac{I_{gas}}{I_{air}} \qquad \text{or} \qquad R = \frac{I_{air}}{I_{gas}}$$

where $I_{gas}$ and $I_{air}$ are the electrical current values after saturation in the presence of reducing/oxidizing gas and synthetic air, respectively. Response and recovery times are determined using the equations:

$$I = \begin{cases} I_0 + A_d + A_g \left( e^{-\frac{t_c}{t_{res}}} - e^{-\frac{t}{t_{res}}} \right), \ I \leq I_0 \\ I_0 + A_d \, e^{-\frac{t-tc}{t_{rec}}}, \ I > I_0 \end{cases}$$

Where $t_{res}$ and $t_{rec}$ are the response and recovery times respectively, $I_0$ is the initial current value, $A_d$ and $A_g$ are the decay and growth amplitudes respectively and $t_c$ is the time where current reaches its maximum value.

## 2.5 Computational methodology

Seven-layer slab models were employed to simulate the $Cs_2AgBiBr_6$ surfaces, in accordance with previous studies.[31,32] These (100) oriented slabs were constructed from a 2×1×1 supercell of the bulk cubic phase of $Cs_2AgBiBr_6$, with lattice constants a= 36.8848Å, b= 11.2499 Å, c= 11.2499 Å, finite in the x-direction with a vacuum gap of 20 Å, and periodic along the y- and z-directions.[33] Specifically, symmetric slabs consisting of 3 (4) $Cs_2Br_2$ and 4 (3) $BiAgBr_4$ alternating layers were constructed for termination A (termination B).

Density Functional Theory (DFT) calculations were conducted using the Vienna ab initio simulation package (VASP).[34,35] The Perdew-Becke-Erzenhof (PBE) parametrization within the Generalized Gradient Approximation (GGA) was employed for the exchange-correlation (XC) functional.[36] The projector augmented wave (PAW) formalism was utilized to model interactions between valence electrons and ions.[37] A plane wave energy cut-off of 400 eV and van der Waals interactions treated with the DFT-D3 scheme of Grimme were considered.[38] Convergence criteria included an energy cut-off of $10^{-6}$ eV for electronic degrees of freedom and a maximum force threshold of 0.01 $eV \cdot Å^{-1}$ for relaxation of atomic positions. Brillouin zone integration was performed using a 1×4×4 Γ-centered k-point mesh, with Gaussian smearing of 0.1 eV. A denser 2×5×5 Γ-centered k-point mesh was used for the PDOS calculations. During structure optimization, the cell volume and shape remained fixed, while atoms in the top three surface layers and the adsorbates were allowed to relax.

The adsorption energies $\Delta E_{ads}$ were calculated as $\Delta E_{ads}=E_{(X+slab)}-E_{slab}-E_X$, where $E_{(X+slab)}$ is the energy of the slab with the adsorbed X molecule (with X being $O_3$, NO, $H_2$, $CO_2$, $CH_4$), $E_{slab}$ is the energy of the slab, and $E_X$ is the energy of the X molecule in vacuum. For the case of NO, the open-shell (spin-polarized) configuration was used.

Charge density difference plots were generated using VESTA software[39] on a 0.001 ($e/bohr^3$) isosurface, calculated as the difference of the electron density between the slab + X complex system and the sum of the isolated components within the conformation of the complex.

# 3 Results and Discussion

## 3.1 Optimal features of $Cs_2AgBiBr_6$ perovskites for $O_3$ detection

Microparticles of the double perovskite $Cs_2AgBiBr_6$, free of capping ligands and exhibiting diverse morphologies and sizes, were synthesized to assess their reactivity towards $O_3$ and determine the optimal characteristics for gas sensing applications. Within the scope of this work, three discrete morphologies of $Cs_2AgBiBr_6$ - sheets, spherical and faceted flower-like particles - were successfully synthesized.

The ligand-free microsheets (Figure 1a), with dimensions of 22.8 ± 6.2 μm, were obtained from the precipitate following a protocol that involved the introduction of the inorganic precursors dissolved in dimethyl sulfoxide (DMSO) into ethanol. Concurrently, the spherical microflowers (Figure 1b), with an average size of 0.90 ± 0.29 nm, were collected from the supernatant of the same mixture. Faceted flower-like microcrystals, sized 2.14 ± 1.29 μm (Figure 1c), emerged upon drop-casting the precursor solution onto a substrate at 80°C. Similar flower-like particles were synthesized by Greul et al. through high-temperature synthesis, and their formation was attributed to the rapid solvent evaporation, which accelerated the crystallization process and resulted in the formation of agglomerates.[40] The first two morphologies were deposited on interdigitated commercial Pt-electrodes for $O_3$ sensing evaluation, while the third was directly grown on the electrodes.

Compositional analysis using Energy-dispersive X-ray spectroscopy (EDS) confirmed the presence of Cs, Ag, Bi and Br in all synthesized morphologies (Figure S1). The crystal structures of the same materials were determined via X-ray diffraction (XRD), revealing that all morphologies were crystalline in the cubic $Cs_2AgBiBr_6$ structure with space group $Fm\bar{3}m$ (ICDD 01-084-8699) (Figure S2a). Notably, additional minor reflections at approximately 30.9° and 44.3° were also detected and attributed to the unreacted precursor AgBr and the secondary phase $Cs_3Bi_2Br_9$, corroborating findings from prior studies on the synthesis of the same material.[40]

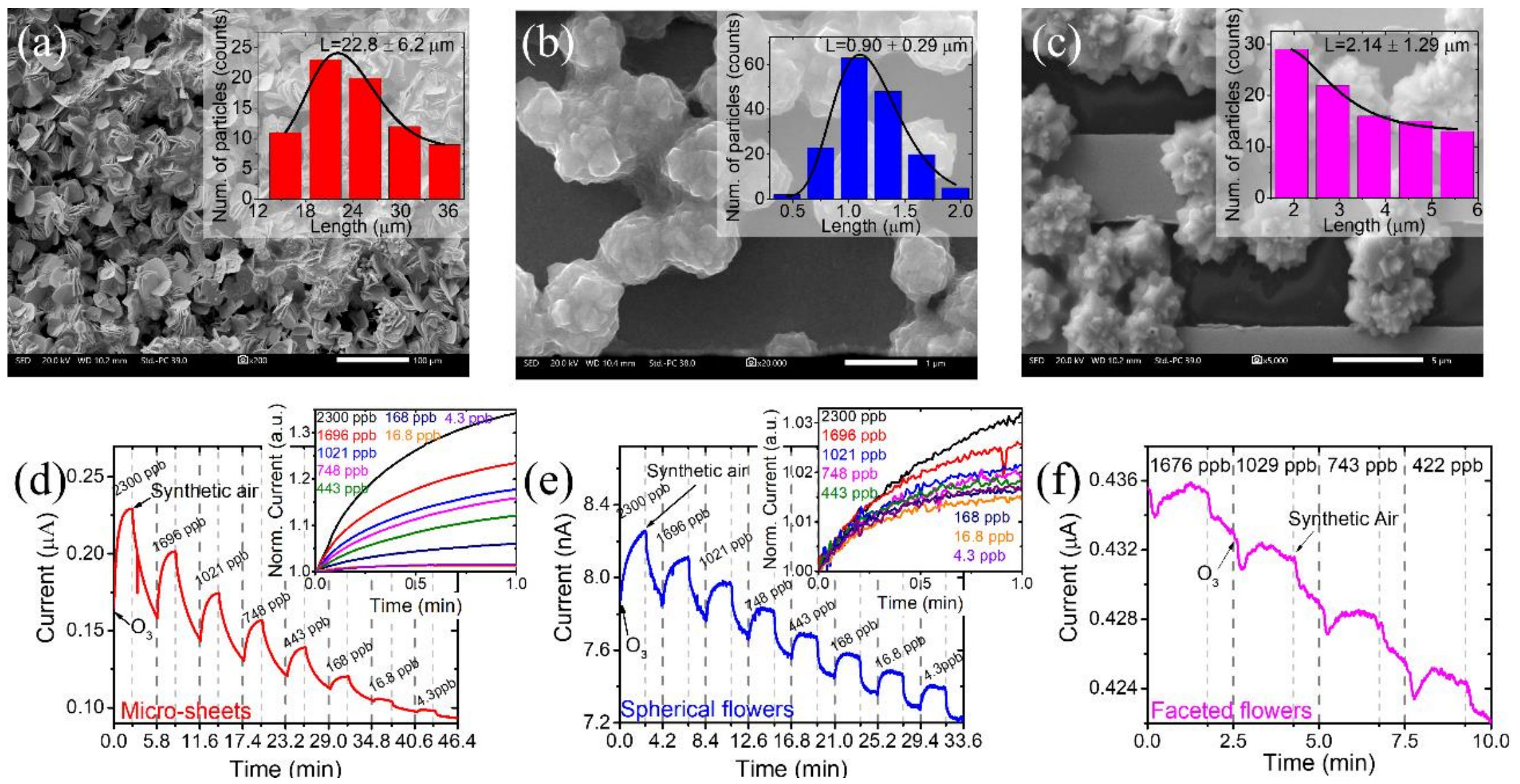

**Figure 1.** FESEM images and size distribution diagrams (insets) of $Cs_2AgBiBr_6$ perovskite-based sensors with microparticle of (a) sheet-, (b) spherical flower- and (c) faceted flower- shaped morphology. (d-f) Electrical response of perovskite-based sensors under different $O_3$ concentrations, at room temperature. Normalized oxidation curves as a function of time (insets in b, c).

The electrical sensing capabilities of the three distinct morphologies of $Cs_2AgBiBr_6$ perovskites were systematically evaluated under standard laboratory conditions, including room temperature and natural ambient light (Figure 1d-f). The experimental setup involved the application of a voltage range from 0.1 to 1 V depending on the morphology, with a constant pressure of 600 mbar. The two larger morphologies operated effectively at a minimal voltage of 0.1 V, whereas this voltage was insufficient to activate the smallest spherical microflowers, likely due to their comparatively wider bandgap (Figure S2b-f). The band gap energy of the spherical microflowers, as determined from the Tauc plot, was found to be 2.30 eV, higher than the 2.05 eV for the microsheets and 2.22 eV for the faceted microflowers. A constant voltage of 1 V was applied throughout the measurement process for the spherical microflowers. Interestingly, a notable power advantage was originated from the ability of the larger particles to operate efficiently at such a reduced voltage. This operational efficiency, combined with their independence from external stimuli such as heat or UV irradiation, significantly reduced the overall energy consumption, offering a more cost-effective solution for the development of portable sensing devices.

Upon exposure to $O_3$, all examined morphologies of the $Cs_2AgBiBr_6$ material exhibited a p-type behavior (Figure 1d-f). This was evidenced by an increase in current intensity upon exposure to $O_3$ until reaching saturation point, and a decrease in current intensity when exposed to synthetic air. This behavior was consistently observed across all tested $O_3$ concentrations, ranging from 2300 ppb to 4.3 ppb. Notably, the microsheet-based sensor demonstrated the ability to differentiate between low $O_3$ concentrations. The normalized oxidation curves highlighted the sensor's capability in detecting $O_3$ levels approximately as low as 160 ppb (Figure 1d, inset), which is well below the 300 ppb OSHA's PEL-STEL limit (Permissible exposure limit – Short-term exposure limit).[41] Conversely, the other samples exhibited poor responses with inability to discern different $O_3$ concentrations, suggesting that both faceted and spherical microflower-based sensors were not suitable for $O_3$ detection, especially at low concentrations (Figure 1e-f). Therefore, the microsheet morphology was identified as the most effective for $O_3$ detection, demonstrating high sensitivity even at lower concentrations. The difference in the sensing behavior between the microcrystals could be possibly attributed to the different total surface but also due to the change in surface termination, as indicated by the different preferential growth orientation observed in the XRD patterns (Figure S2a). This finding was also contrasted with a report on analogous CuO morphologies by Luo et al., where flower-like configurations showed enhanced responses to volatile gases compared to their sheet-like counterparts.[42] This difference can be attributed to porosity variations, defect density or even defect type.[43–47] Notably, small nanocrystals were presented on the surface of the microsheet-based sensor (Figure S3). Similar small particles were observed in the case of the flower-like structures of the higher responsiveness.[48]

## 3.2 Response to $O_3$ and stability to time, humidity and temperature of the microsheet-based sensor

**Sensor's performance.** The sensor's response, calculated from the $I_{gas}/I_{air}$ ratio (where $I_{gas}$ and $I_{air}$ denoted the maximum and minimum current values, respectively, during oxidation), exhibited a distinct correlation with $O_3$ concentration, increasing from approximately 1.07 at 168 ppb to 1.38 at 2300 ppb (Figure 2a). Notably, the response could be fitted into two distinct linear curves and particularly, it could be observed that in concentrations lower than approximately 750 ppb, the response demonstrated an increasing rate, as indicated by the larger slope of the curve (blue fitting curve, Figure 2a). Conversely, at higher concentrations, the response displayed a slower rate of increase (green fitting curve, Figure 2a). A similar transition from high to low response rate was reported in prior studies.[49, 50] Sun et al. observed this phenomenon in multi-walled carbon nanotubes upon exposure to $O_3$,[49] while Cui et al. reported a similar occurrence in 2D phosphorene nanosheets when subjected to $NO_2$.[50] In the former case, the shift in response trend was attributed to the saturation of the perovskite surface with $O_3$ molecules while in the latter, it was linked to the decrease in the mobility of charge carriers due to scattering effects from gas molecules adhering to the sensor's surface.

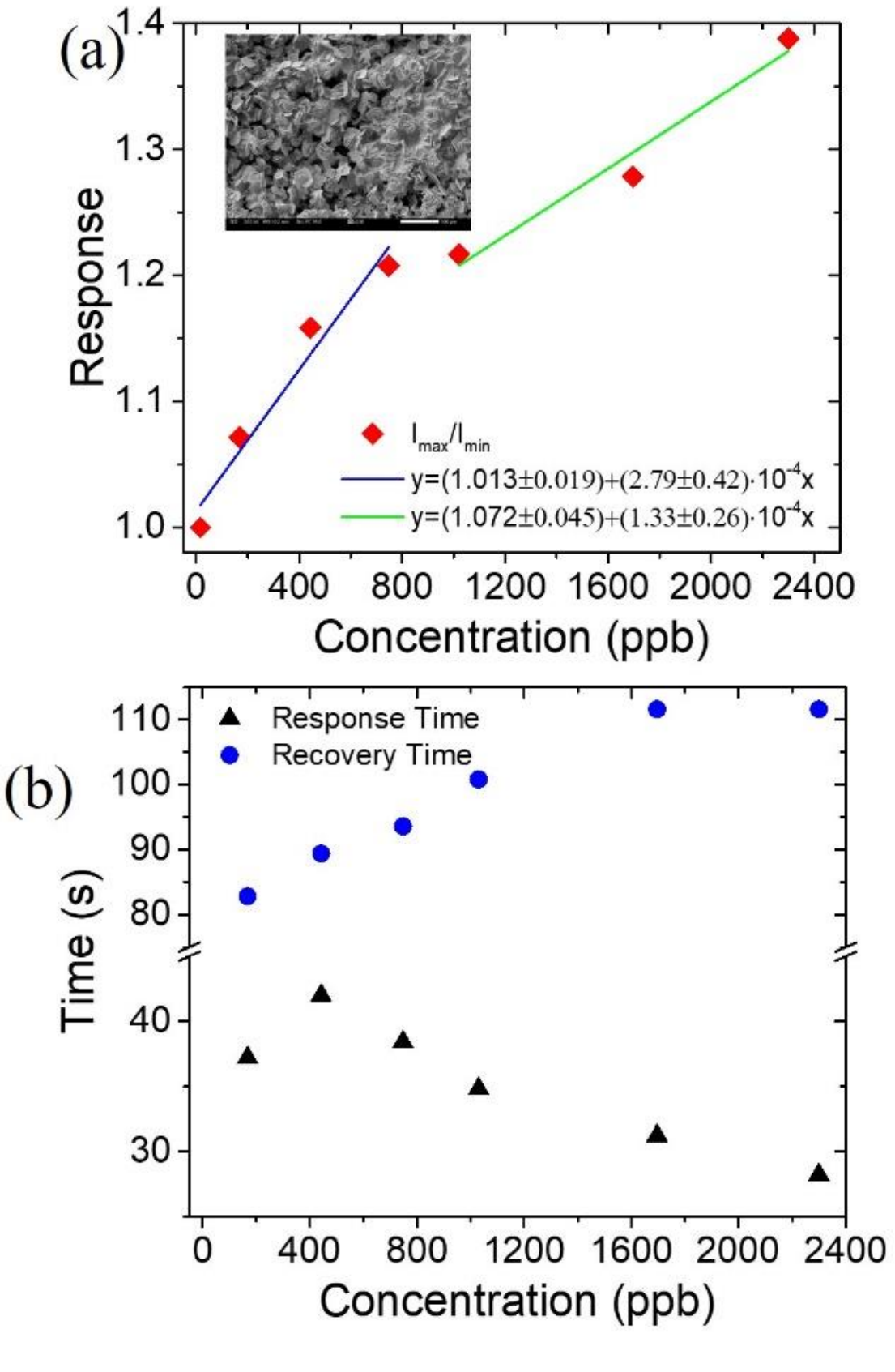

**Figure 2.** (a) Calculated response and (b) Response/Recovery times of the sensor as a function of $O_3$ concentration.

Achieving a rapid response, fast gas detection, and quick recovery times is essential for a high-quality sensor. The temporal metrics for response ($t_{res}$) and recovery ($t_{rec}$) times in relation to $O_3$ concentration are critical parameters. These are determined by analyzing the exponential growth and decay characteristics of the sensor's electrical response across a range of gas concentrations.[51] It is noteworthy that the microsheet-based sensor demonstrated a rapid response, with response times of less than 30 seconds for elevated gas concentrations and under 50 seconds for lower concentrations (Figure 2b). Furthermore, the sensor's recovery process was markedly efficient, with durations of less than 2 minutes. At the detection threshold of 168 ppb, the sensor is capable of recovering in a period of less than 83 seconds.

Upon detailed comparison with leading sensors operating at room temperature, primarily those utilizing metal oxides synthesized through high-temperature or lengthy processes, the microsheet-based sensor's performance is comparable to certain sensors at similar $O_3$ concentrations (Figure 3, orange frame, Table S1). These sensors include $AgIn_2O_3$ (1wt%),[52] NiAl,[53] and Au@$TiO_2$,[54] within the $O_3$ concentration range of 100 to 500 ppb and operate without requiring external stimuli. However, some sensors, such as ZnO-$SnO_2$[55] and $In_2O_3$[56], require irradiation to function effectively. Conversely, our sensor exhibits a reduced response compared to others, such as ZnO,[57] InGaZnO (IGZO),[58] ZnO-gold,[57] 0.1% Au $TiO_2$-$WO_3$,[59] and $TiO_2$-$In_2O_3$[60]. The operation of these sensing elements is triggered by an external stimulus, such as UV or LED irradiation, leading to increased cost and higher power consumption. Notably, carbon nanotubes, a different type of sensing element from metal oxides, exhibited a similar response pattern without external triggering, even when tested at significantly higher $O_3$ levels (5000 ppb).[49]

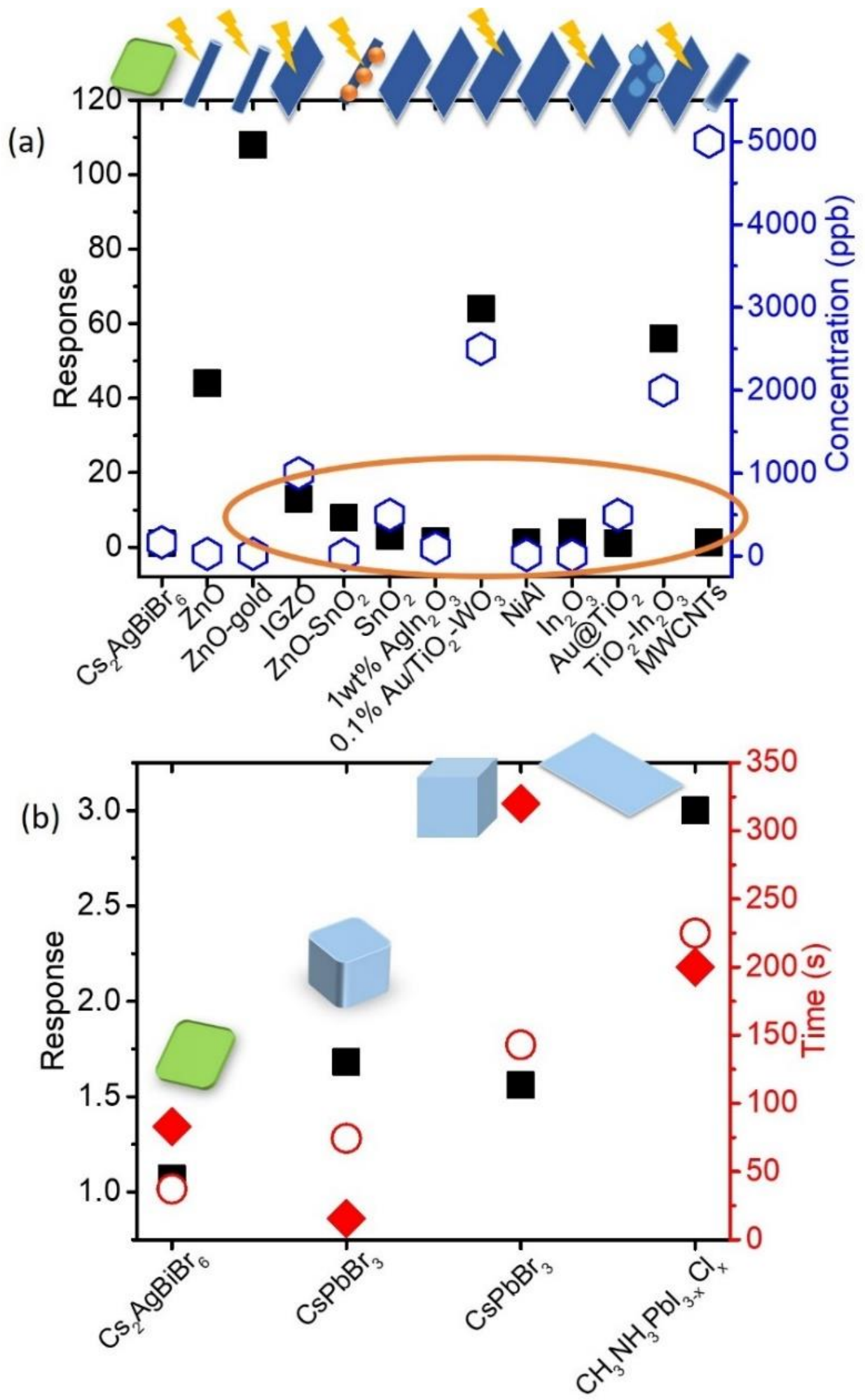

**Figure 3.** (a) Response of the microsheet-based sensor to $O_3$, in comparison with other semiconducting sensing elements that have been documented in the literature. [57,58,55,61,52,59,53,56,54,60,49] These comparisons are made at the respective detection limits of the sensors, noting whether they operate at room temperature and if light irradiation is applied. The orange circle highlights the materials with response characteristics similar to $Cs_2AgBiBr_6$ sensor. (b) Response at 180 ppb of $O_3$ and the respective response/recovery times in relation to metal halide perovskites previously reported in literature .[51,62,63] The response time is indicated by open symbols and the recovery with filled one. The $Cs_2AgBiBr_6$ microsheets are highlighted in green. Additionally, the diagram provides a visual representation of the materials' morphology and indicates whether an external stimulus is utilized during their operation.

Promising results have emerged regarding the performance of the $Cs_2AgBiBr_6$ microsheet-based sensor compared to other metal halide perovskites utilized for $O_3$ detection at room temperature without requiring external triggering (Figure 3b, Table S2). While the response of Pb-free perovskite-based sensor may not be the highest among all the metal halide-based sensors at comparable $O_3$ concentrations (180 ppb), its response time is notably the shortest, and its recovery time ranks among the quickest. Despite not having the highest response, this cost-effective, environmentally and health-friendly alternative, with its sheet-like morphology, can efficiently detect low gas concentrations within a few seconds without the need for any external light source. This underscores the potential of $Cs_2AgBiBr_6$ microcrystals, which could be further optimized, akin to their well-established counterparts in the field.

**Short- and long-term stability**. The short-term repeatability and long-term stability of the sensors constitute critical parameters for ensuring reliability and accuracy over time. Each sensing cycle exhibited a consistent pattern characterized by a current increase in the presence of the oxidizing gas, followed by a fully recovery in the presence of synthetic air, indicating the good reversibility of the sensor (Figure 4a). The absence of drift, even after three successive cycles, further suggests that the sensor can be utilized for continuous air monitoring. In addition, the long-term stability of the sensor was evaluated after six weeks, showing similar sensing behavior, reinforcing the sensor's reliability, (Figure 4b) while both morphological and structural characteristics remained unaltered at the end of the sensing process over time (Figure S4a-b). The chemical stability of $Cs_2AgBiBr_6$ micro-sheets before and after $O_3$ exposure was investigated by X-ray photoelectron spectroscopy (XPS). Particularly, narrow scans were conducted for all the relevant elements of interest, including Cs 3d, Ag 3d, Bi 4f and Br 3d core levels, however, no chemical shifts were observed, suggesting that the chemical environment remained unaffected by the $O_3$ treatment (Figure S5).

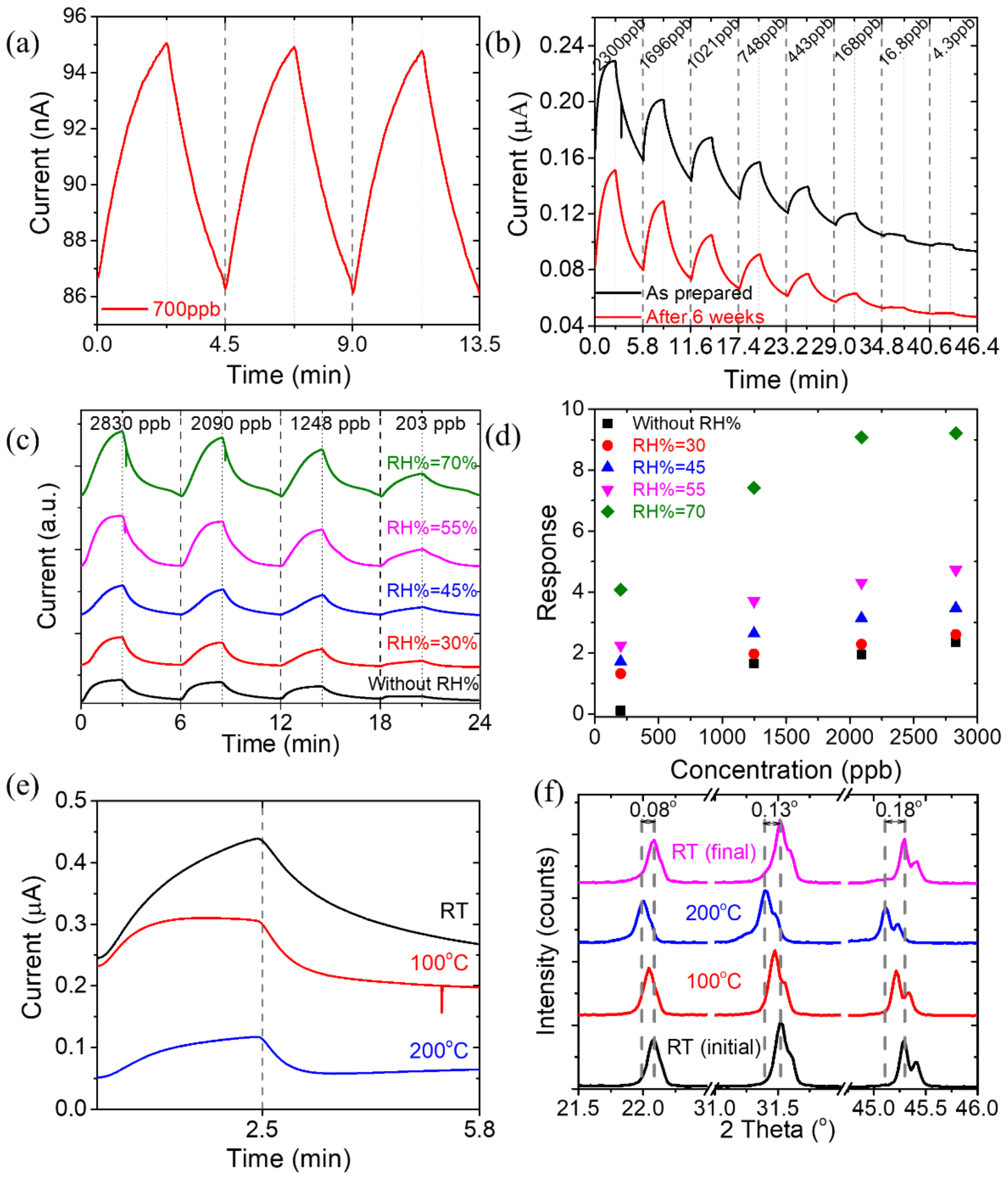

**Figure 4.** (a) Short-term stability diagram of $Cs_2AgBiBr_6$ micro-sheets upon three successive cycles. (b) Long-term stability diagram of the as prepared and after six weeks upon exposure to different $O_3$ concentrations. (c) Electrical response over time without and with relative humidity in the range of 30% to 70% under different $O_3$ concentrations. (d) Calculated response as a function of concentration for different relative humidity levels. (e) Temperature dependence of the electrical response at 2000 ppb $O_3$ exposure. (f) Temperature dependent XRD pattern.

**Stability against humidity.** In various real-world applications that involve semiconductor gas sensors, including air-quality monitoring or agricultural sensing systems, the effect of humidity on sensor performance remains a critical consideration. To probe the effect of humidity, the stability of the microsheet-based sensor was evaluated under the influence of controlled relative humidity (RH%) levels at room temperature under different $O_3$ concentrations ranging from 2890 down to 168 ppb (Figure 4c). The gas sensing tests were conducted under both dry conditions and at controlled relative humidity levels with RH% from 30% to 70%. The results showed similar electrical current diagrams across all humidity conditions, proving the sensor's ability to detect $O_3$ effectively in such conditions. Notably, the sensor's responsiveness improved significantly with increasing RH%, especially at lower $O_3$ levels, where the response rate increased by more than tenfold (Figure 4d), while it retained its morphological features (Figure S4e). This behavior stands in contrast to most chemiresistive gas sensors that their response typically deteriorates under high humidity levels,[63,64] making the microsheet-based sensor even more noteworthy.

A plausible explanation for the enhanced gas sensing behavior upon humidity could be assigned to the adsorption of water molecules on the surface of the perovskite material. Similar to $Cs_2SnCl_6$, $Cs_2TeCl_6$ and $Cs_2AgBiBr_6$ perovskites that have been studied as effective humidity sensors, the adsorption of water molecules on the surface of these materials likely enhances the sensor's conductivity.[65,66,67] In particular, the adsorption of humidity involves both chemisorption and physisorption, occurring at low and high humidity levels, respectively. At low humidity conditions, the $H_2O$ molecules are chemically adsorbed onto the perovskite surface, releasing a proton ($H^+$). The $H^+$ is then transferred to an adjacent water molecule, forming $H_3O^+$. Subsequently, the proton undergoes a series of transfers from site to site, which leads to an increase in the sensor's conductivity. On the other hand, under high humidity environment, water molecules form multiple physisorbed layers on perovskite' surface, creating a liquid-like network of hydrogen-bonded water molecule layers. The hydration of $H_3O^+$ is energetically favored in the presence of liquid water, resulting in proton-transfer described by the equation $H_2O + H_3O^+ = H_3O^+ + H_2O$. The hydrated proton then moves between adjacent $H_2O$ molecules within a continuous layer of water, contributing further in the increase of conductivity.[68]

The effect of humidity on the $Cs_2AgBiBr_6$ perovskite structure at the end of the sensing process was investigated by XRD (**Figure S4c and d**). The analysis revealed a slight shift of approximately 0.02° in the characteristic peaks of $Cs_2AgBiBr_6$ towards higher angles compared to the initial pattern. This shift indicates a compressive strain[69], possibly arising from the humidity-induced formation of crystallographic defects. Additionally, the appearance of a distinct diffraction peak at approximately 32.3° is likely attributed to the presence of tetragonal BiOBr crystal phase, suggesting a partial decomposition of the material into $BiBr_3$, followed by its hydrolysis into BiOBr.[70,71]

**Stability against temperature.** To further investigate the stability of the sensor under harsh conditions, the effect of working temperature was evaluated. Subsequent experiments were conducted at variable temperatures, specifically under the exposure to $O_3$ at a concentration of 2000 ppb (Figure 4e). Interestingly, the sensor was able to operate even at elevated temperatures, suggesting its stability in such conditions and its potential applications in environments with fluctuating or high temperatures. However, it was noticed that by increasing the temperature, the electrical current decreases from approximately 0.25 μA down to 0.05 μA (Figure 4e). This phenomenon could be explained by the competition between thermal adsorption and desorption of atmospheric oxygen on the perovskite surface. A similar explanation was given by Bejaoui et al for the CuO $O_3$ sensor.[72] This study suggested that at room temperature, the conduction process was dominated by low resistance due to the high density of holes. However, when the temperature was increased, the desorption was increased, leading to a decrease in the hole accumulation layer and consequently to the increase of the resistance.

Furthermore, to elucidate any alterations induced by the thermal process to the perovskite material, which could potentially explain the observed sensing curves, temperature-dependent XRD analysis was performed (Figure 4f). Intriguingly, no new diffraction peaks emerged under high-temperature conditions; however, significant shifts toward lower angles were noted. These shifts indicated the lattice expansion of $Cs_2AgBiBr_6$ at elevated temperatures (100 and 200°C), suggesting a modification in lattice parameters that could potentially account for the differences in conductivity. Upon cooling to room temperature, the diffraction peaks reverted to their original positions, implying that these changes were reversible. Additionally, to assess the synergistic effect of $O_3$ together with the temperature on the material, XRD patterns were acquired at the conclusion of the sensing process following exposure to $O_3$ (Figure S6). These patterns remained unaffected, confirming that the crystal structure was preserved at the end of the sensing process, with side phase intensities diminishing at 200°C, consistent with the known effect of annealing at 250°C for the complete conversion of the precursors into the desired $Cs_2AgBiBr_6$.[40]

### 3.3 Response and selectivity in various gases

To ascertain the selectivity of the $Cs_2AgBiBr_6$ microsheet-based sensor towards $O_3$ gas, additional tests with both reducing and oxidizing gases were conducted, including nitric oxide (NO), hydrogen ($H_2$), methane ($CH_4$), and carbon dioxide ($CO_2$) (Figure 5 and S7). Similarly to $O_3$, in the presence of NO, the sensor exhibited a p-type sensing behavior, however, to prevent the formation of $NO_2$, pure $N_2$ was employed as recovery gas (Figure 5a). Remarkably, the NO sensor demonstrated the ability to resolve low gas concentrations, up to 2 ppm (Figure S8a), with excellent stability even after multiple sensing cycles (Figure 5a inset). However, the sensor's response to $Cs_2AgBiBr_6$ microsheets was modest, even at 10 ppm (Figure 5b). Moreover, contrary to $O_3$ response, which exhibits a steeper slope at lower concentrations and a more gradual increase at higher concentrations, the NO response shows a smaller slope initially and a more pronounced increase at higher concentrations. This behavior could indicate that gas molecules might need to reach a certain concentration threshold before significantly affecting the sensor.

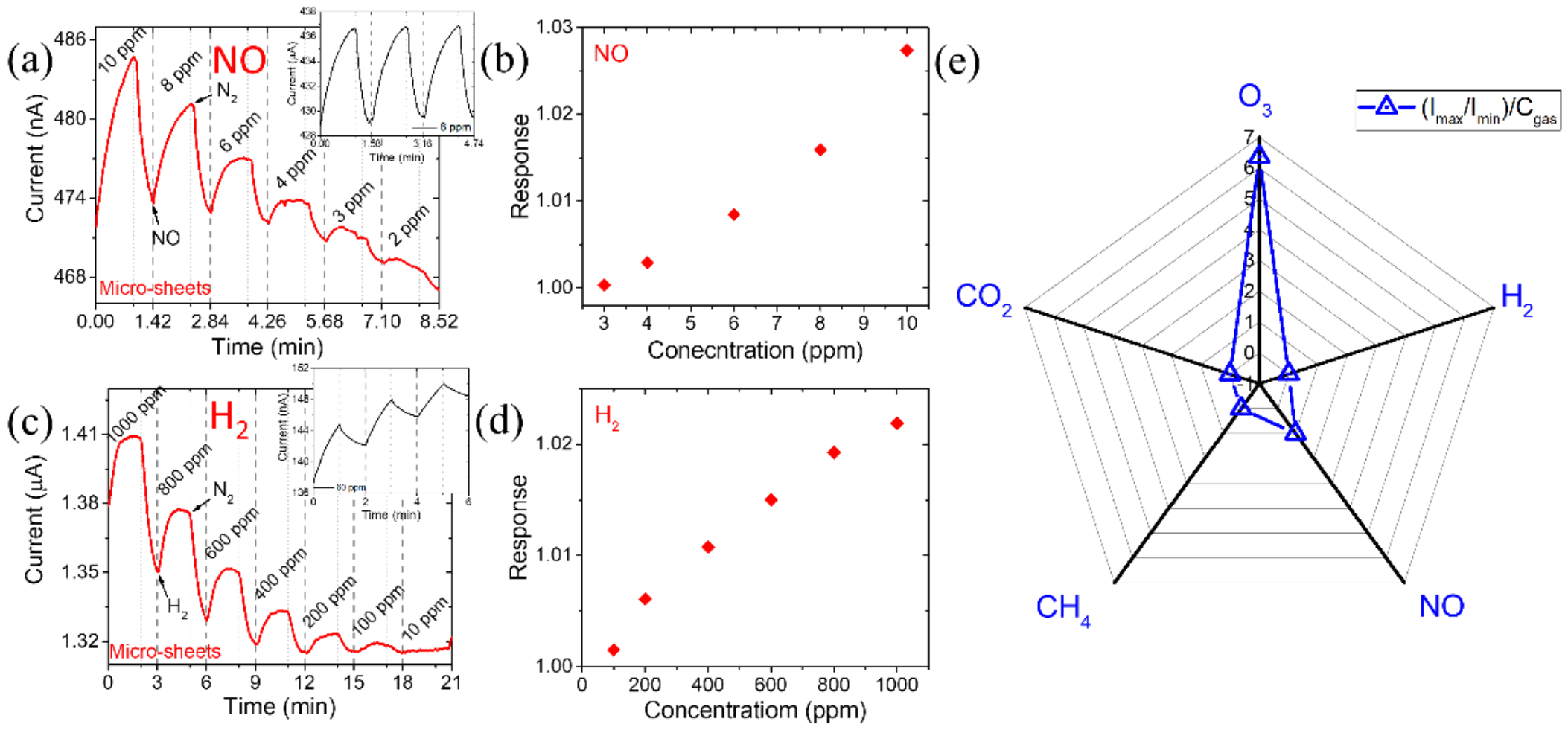

**Figure 5.** Electrical response of the microsheet-based sensor when detecting (a) NO and (c) $H_2$ under different gas concentrations, at room temperature. Calculated response as a function of (b) NO and (d) $H_2$ concentration. The insets in (a) and (c) show the normalized oxidation curves of NO and $H_2$, respectively,

as a function of time. (e) Radar chart showing the normalized response ($I_{gas}/I_{air})/C_{gas}$) of the sensor in the different gases.

Unlike $O_3$ and NO, $H_2$ is not considered toxic, however, in concentrations higher than 40,000 ppm (in air) it can pose a high explosion risk.[73] When exposed to $H_2$, the sensor's current diagram was akin to that of NO, despite $H_2$'s reducing properties, suggesting a distinct sensing mechanism. It is also worth noting that, while $Cs_2AgBiBr_6$ showed remarkable repeatability under both $O_3$ and NO, the $H_2$ sensor showed a significant baseline drift, indicating a possibly irreversible sensing process. This drift could be potentially attributed to $H_2$ molecules diffusing into the perovskite's crystal lattice, possibly inducing changes in the material.[51] Studies suggest that hydrogen atoms preferentially adsorb onto the interstitial sites of the $Cs_2AgBiBr_6$ lattice, creating new bonding states.[74] Despite the sensor's ability to detect $H_2$ at levels as low as 20 ppm, its overall response proved to be poor, even at higher concentrations.

In order to examine the effect of $H_2$ on $Cs_2AgBiBr_6$ microsheets at the end of the $H_2$ sensing process, XPS was employed. Narrow scans of Cs 3d, Ag 3d, Bi 4f, and Br 3d of the microsheets before and after $H_2$ exposure showed, similarly to $O_3$, no chemical shift, indicating the absence of chemical interactions on the surface of the material (Figure S9a-d), while the survey XPS spectra showed no alterations as well (Figure S9e). The crystal structure upon $H_2$ exposure was also evaluated by the means of XRD and a noticeable increase in the peak intensity of (311) plane was observed, suggesting structural modifications upon $H_2$ exposure (Figure S9f). Additionally, the peak intensities of the byproducts were slightly increased, indicating a partial decomposition of the perovskite material upon $H_2$ exposure. These findings suggested structural changes and possible partial decomposition of the perovskite material due to $H_2$ exposure.

Finally, in the context of reducing gases, $CO_2$ and $CH_4$ were also tested for cross-sensitivity (Figure S7). Contrary to $H_2$, both sensors exhibited an n-type sensing behavior, as expected, with poor current changes under high gas concentrations, indicating low charge transfer between the perovskite material and the $CO_2$ and $CH_4$ gases at room temperature.

Considering the observed variations in current corresponding to the concentrations of each target gas, a radar chart delineating the normalized response versus gas concentration was constructed to evaluate the selectivity of the developed sensor (Figure 5e). The graphical representation definitely demonstrated the sensor's pronounced selectivity towards $O_3$, surpassing its sensitivity to other tested gases. Notably, NO and $H_2$ caused moderate responses, while $CO_2$ and $CH_4$ exhibited the minimal sensor response.

### 3.4 DFT calculations to predict $Cs_2AgBiBr_6$ gas sensor selectivity

To shed light on the response trend to the different tested gases, first-principles calculations were employed to investigate the adsorption interactions between the $Cs_2AgBiBr_6$ surface and $O_3$, NO, $H_2$, $CO_2$ and $CH_4$ target molecules. Since an ideal sensor material is characterized by its ability to form robust bonds with the target gas while inducing significant alterations in its electronic structure upon adsorption, investigation of the binding strength of the target molecules to the perovskite surface as well as of the density of states through Density Functional Theory (DFT) simulations serves as a promising avenue to elucidate the experimental observations. However, finding a suitable surface model to accurately simulate the experimental conditions and provide meaningful insights is challenging due to several reasons. Perovskite surfaces can exhibit structural complexities such as different terminations, defects and reconstructions, which pose difficulties in selecting the most relevant to the experiment surface model. Introducing defects or dopants into the surface model adds another layer of complexity, as does determining the energetically favorable configuration of the adsorbate molecules on the non-ideal surface. In addition, balancing computational feasibility with the need for accurate representation of the surface properties further complicates the selection, as the size of the simulation cell needs to be large enough to capture the essential surface properties while remaining small enough to allow for an extensive investigation of the configurational space.

Herein, the (100) oriented surfaces based on the bulk cubic crystal structure of $Cs_2AgBiBr_6$, as synthesized by A. H. Slavney et al., were selected .[33] Taking into consideration insights from prior theoretical investigations,[31,32] these surfaces were modeled as seven-layer slabs of alternating $BiAgBr_4$ and $Cs_2Br_2$ layers. Two possible terminations exist for these models; termination A (TA: BiBr/$AgBr_3$) and termination B (TB: CsBr), both illustrated in Figure S10 along with the different adsorption site types for each one. Prior theoretical research by G. Volonakis and F. Giustino demonstrated that under various synthetic conditions, CsBr-terminated (TB) surfaces are favored in $Cs_2AgBiBr_6$ perovskites.[31] However, our sensor material may comprise a combination of both terminations, highlighting the importance of examining both cases to assess the experimental sensing trends. For both TA and TB ideal (i.e., non-defected) surfaces, extensive research was conducted to identify the energetically preferable adsorption site for each target molecule, with overall results shown in Figure 6 and Figure S11. Notably, $O_3$ showed the strongest interaction (-0.7 eV) regardless of the termination, whereas NO exhibited a strikingly different behavior with respect to the termination of the surface; while approaching the TA surface yielded an energetically favorable conformation of -0.17 eV, the TB surface showed almost no interaction with NO (0.01 eV). The experimentally observed significant response of the sensor to NO gas cannot be simply extracted from adsorption energy alone, and is very likely due to some long-range effect that is inaccessible by DFT calculations, or the presence of defects on the surface. Vacancies, interstitials and dopants are among the common defects in these materials. Such defects may have significant implications on the material's properties and performance, affecting its electronic behavior. In fact, the stable chemical potential range for pristine (i.e., defect-free) $Cs_2AgBiBr_6$ is quite narrow, as demonstrated in the theoretical study of Z. Xiao et al..[75] Amongst others, Bi vacancies and AgBi antisites are deep acceptors and the dominant defects under Br-rich conditions, Ag vacancies are shallow acceptors and can easily form, and Br vacancies are identified as intrinsic donors. Since Br vacancies ($V_{Br}$) can be found in both the TA and TB uppermost layer of our surface model,[76] the role of this type of defects on the performance of the developed sensor was selected to be further investigated.

Focusing on the CsBr-terminated surfaces (TB), two different sites for Br atoms exist, thus two different types of Br vacancies; Br atoms that bond to Bi atoms (Br-Bi) and Br atoms that bond to Ag atoms (Br-Ag). According to P. Chen et al.[32], Br atoms are more easily removed from (Br-Bi) sites, due to lower formation energy. Following this result,1 Br (Br-Bi) atom was removed from the uppermost layer of the TB surface. The effect of the $V_{Br}$ (Br-Bi) on the adsorption energies and configurations of the target molecules compared to the defect-free case is illustrated in Figure 6. The results of the $V_{Br}$ (Br-Ag) sites of TB surfaces are presented for comparison in Figure S11. In the most astonishing cases of $O_3$ and NO, once a $V_{Br}$ (Br-Bi) is introduced to the perovskite surface, $O_3$ enters the vacancy site, adopting a bidentate conformation and resulting in a more than four times stronger interaction (-3.23 eV) compared to the defect-free surface. At the same site type, the NO molecule showed an even more remarkable hundredfold enhancement (-1.26 eV) compared to the non-interacting defect-free surface.

Additionally, in the presence of $H_2$, the interaction between molecular hydrogen and the perovskite surface is minimal, with a nearly negligible binding energy (~ -0.1 eV) across various non-defected and defected surface scenarios, as indicated in Figures 6 and S12. However, while $H_2$ molecules do not exhibit strong interaction with the perovskite surface, dissociation of $H_2$ into two hydrogen atoms (H*) occurs readily on the defected surface (TB surface with $V_{Br}$ (Br-Bi)), releasing 0.8 eV of energy. This dissociation process is not favorable in ideal, non-defected surfaces, suggesting the presence of vacancies in the experimentally synthesized sensor.

The interaction strength was also enhanced in the cases of $CO_2$ and $CH_4$ (Figure S11). For $CO_2$ the interaction almost doubled and the molecule adopted a bended configuration within the vacancy site. On the other hand, for $CH_4$ there was still no indication of significant interaction, neither from the binding energy values nor from the adopted configuration upon adsorption. It should be noted that binding energies higher than ~ -0.3 eV do not suggest good selectivity, as $N_2$ is found to bind with -0.27 eV on defected TB surfaces with $V_{Br}$ (Br-Bi) (Figure S13).

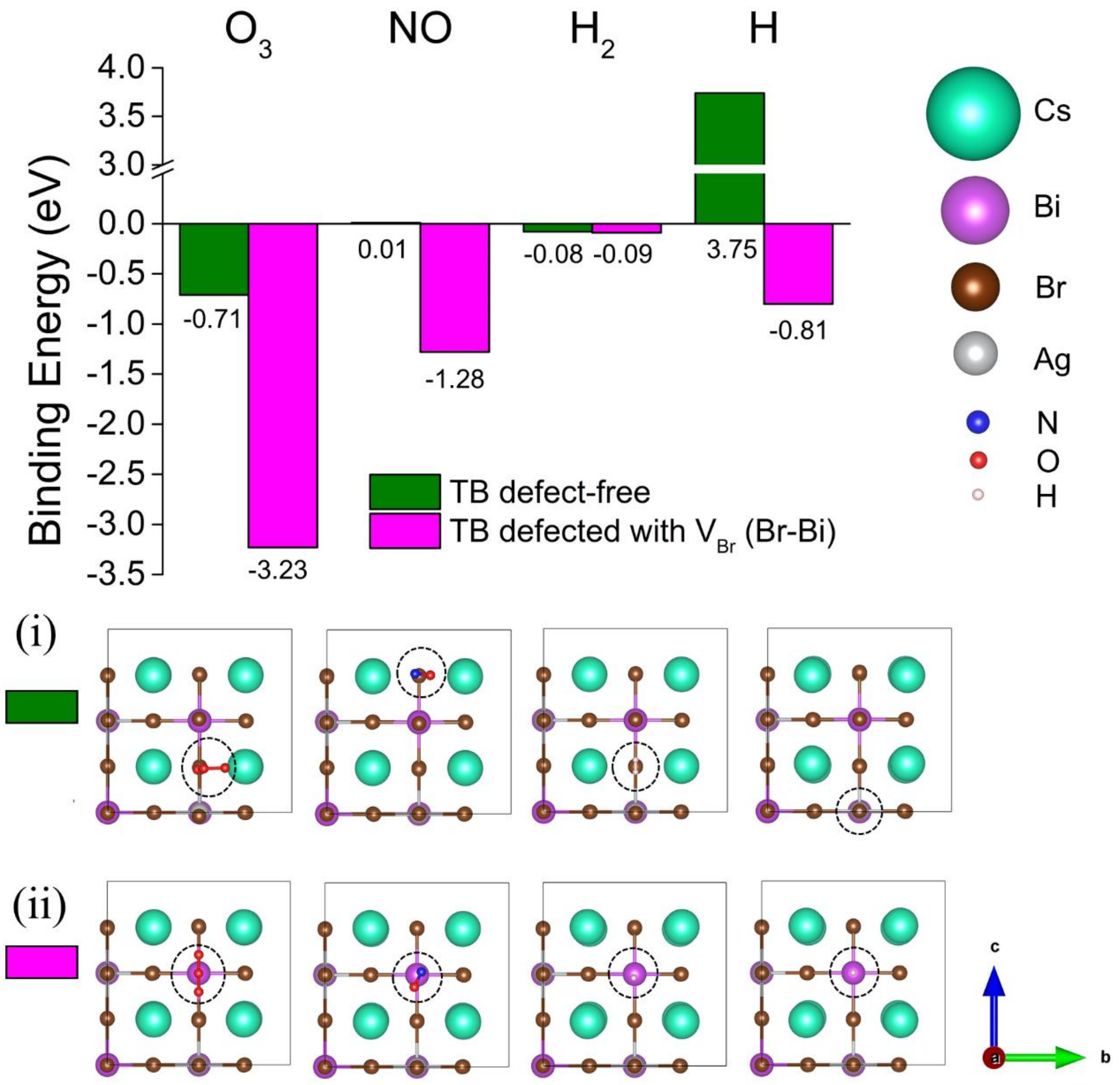

**Figure 6**. Adsorption energies in the strongest binding site for $O_3$, NO, $H_2$ and H target species, on the TB surface without defects (green) and on the defected TB with $V_{Br}$ (Br-Bi) surface (magenta). (i) The strongest binding sites of each target species ($O_3$, NO, $H_2$ and H, respectively) on the TB surface without defects and (ii) on the defected TB surface with $V_{Br}$ (Br-Bi). Cs, Ag, Bi, Br, O, N, and H atoms are illustrated as cyan, silver, purple, brown, red, blue, and white spheres, respectively.

To elucidate the underlying electronic structure changes upon adsorption of the target gases, the Partial Density of States (PDOS) was calculated at the global minimum conformation for each gas case (Figures 7 and S14). In the PDOS of the ideal, defect-free surface without adsorbed gas (Figure 7a),the valence band (VB) and conduction band (CB) are located at approximately -0.5 eV and -1.0 eV, respectively, with their orbital contributions consistent with those of bulk $Cs_2AgBiBr_6$ as calculated by E.T. McClure et al..[77] Once a Br atom that is connected to a Bi atom is removed from the uppermost layer of the surface, i.e., when a $V_{Br}$ (Br-Bi) is introduced to the surface, both CB and VB are shifted to lower energy values, and new states from the Br and Bi atoms emerge near the Fermi level, narrowing the electronic gap (Figure 7b), a result aligned with the findings of P. Chen et al.[32] These defect-induced states vanish upon the adsorption of strongly interacting gases like $O_3$, NO, and H, which fill the $V_{Br}$ site (Figure 7c, d, and f), while persist in the case of the weakly interacting $H_2$ molecule (Figure 7e). Noteworthy, $O_3$ and H adsorption (Figure 7c and f), is accompanied by an observed "healing" effect and opening of the gap, with H returning the PDOS profile to a non-defected state, as most hydrogen states are located deeper in the VB. In the case of $O_3$, new states from the oxygen atoms emerge at the VB, with a prominent peak at the valence band maximum (VBM). A different effect is observed when NO molecules occupy the vacancy sites as new distinct states appear around the Fermi level, originating primarily from the N and O atoms with secondary contributions from the Bi and Br atoms (Figure 7d).

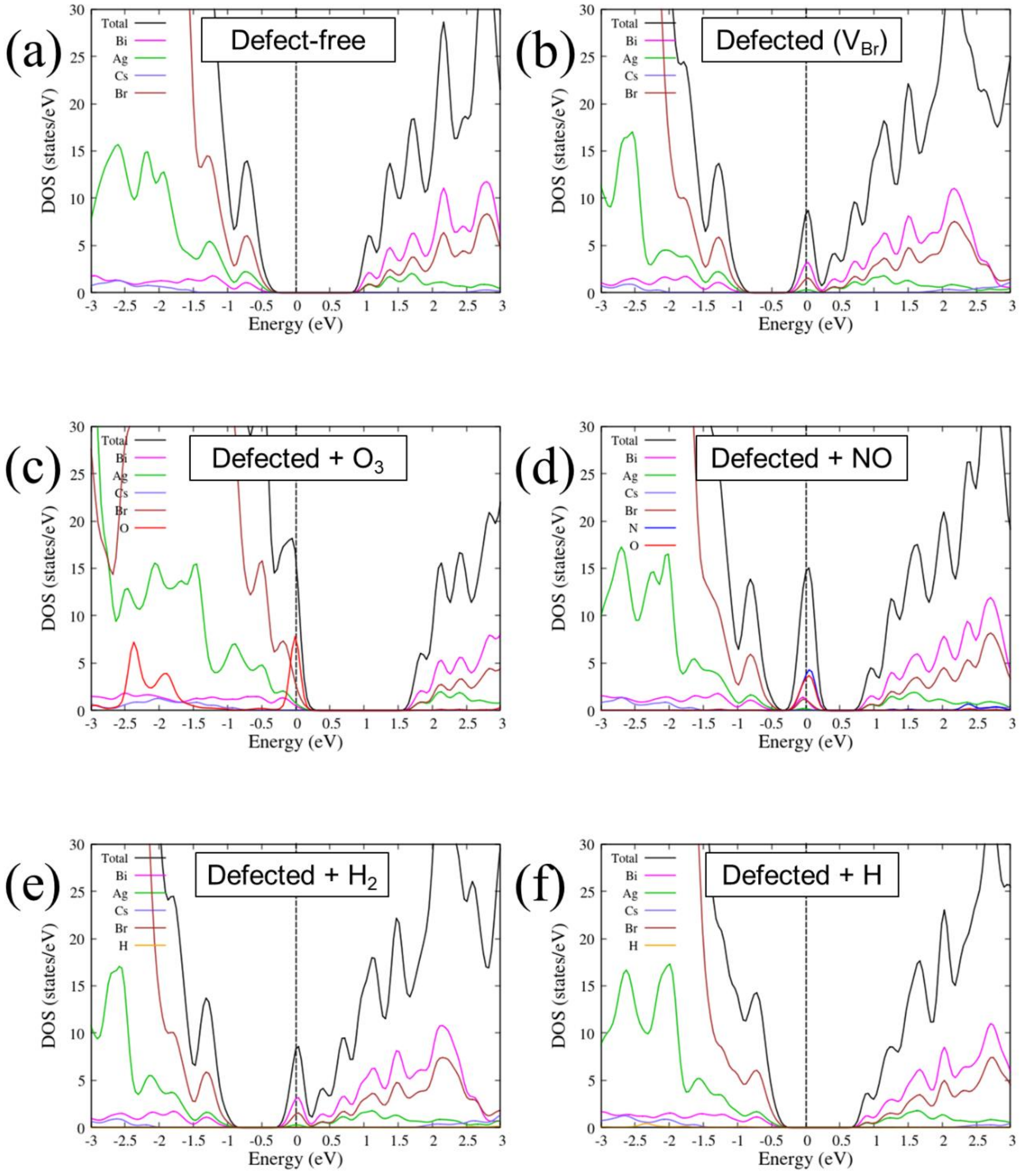

**Figure 7.** Partial density of states (PDOS) of the (100) oriented TB $Cs_2AgBiBr_6$ model surfaces (a) without defects and (b) defected with $V_{Br}$ (Br-Bi). PDOS of the defected with $V_{Br}$ (Br-Bi) TB surface upon adsorption of (c) $O_3$, (d) NO, (e) $H_2$, and (f) H.

A complementary perspective into the electronic structure changes upon gas adsorption is provided also in Figure S15, where the charge density difference plots for all target molecules with the TB defected $V_{Br}$ (Br-Bi) surface are presented, showing significant charge transfer from the uppermost surface atoms to the strongly absorbed species.

## 4. Conclusions

This research marks the first use of Pb-free, environmentally-friendly $Cs_2AgBiBr_6$ double perovskites for $O_3$ detection down to a few hundred ppb without using external stimuli such as heating or irradiation. Furthermore, it investigates the $O_3$ sensing ability of three ligand-free distinct shapes –microsheets, spherical and faceted microflowers– of $Cs_2AgBiBr_6$ double perovskites revealing that the choice of shape significantly influences the gas sensing properties of $Cs_2AgBiBr_6$ perovskites. Among the synthesized morphologies, the microsheet-based sensor proved to be the most effective morphology for $O_3$ detection, capable of resolving concentrations down to 168 ppb under a very low power (0.1V), with a response time less than 40 seconds. This performance surpasses that of traditional toxic lead-based metal halide perovskite counterparts employed for $O_3$ detection to date that operates at room temperature and comparable to the most of state-of-the-art materials used $O_3$ detection operating at room temperature as well. Additionally, the microsheets displayed good reversibility after repetitive gas sensing cycles and remarkable stability over time. Unlike most chemiresistive metal oxide-based sensors, $Cs_2AgBiBr_6$ microsheets were capable to operate under harsh conditions, including high relative humidity levels and elevated temperatures.

Moreover, the cross-sensitivity of the microsheet-based sensor towards other gases, including NO, $H_2$, $CO_2$, and $CH_4$, demonstrates an excellent sensitivity of the perovskite sensor for $O_3$ while also showing great potential for detecting NO and $H_2$. However, it proves unsuitable for detecting $CO_2$ and $CH_4$ due to insufficient signal generation. Theoretical insights evaluating the relative trends of adsorption energies for the target gases, along with the emerging characteristics in the DOS profile upon their adsorption, further highlighted the crucial role of surface defects in the sensing process, with Br vacancies identified as the most effective active sites for $O_3$ detection.

Overall, this work contributes to the development of room-temperature, low-powered, environmentally-friendly gas sensing devices, expanding the understanding of lead-free metal halide perovskites for gas detection. The combination of experimental and computational approaches provides valuable insights into the active sites and sensing mechanisms of metal halide perovskites, paving the way for further research and applications in air quality monitoring.

## Supporting Information

EDS analysis, XRD pattern and absorbance curve of the microcrystals of the three morphologies, FESEM images of $Cs_2AgBiBr_6$ micro-sheets in high resolution. Tables that compare the microsheet-based sensor with other perovskites and non-perovskite-based sensors which operate at room temperature. Characterization of the microcrystals after the sensing process with SEM, XRD, XPS. XRD after annealing. Electrical response under $CO_2$ and $CH_4$. Normalized curves upon different gases. DFT calculations: absorption, density of states and charge density difference plots

## Acknowledgements

This research project has received funding from the EU's Horizon Europe framework programme for research and innovation under grant agreement BRIDGE (n. 101079421 from 01/10/2022 – 30/9/2025). In addition, FLAG-ERA Joint Transnational Call 2019 for transnational research projects in synergy with the two FET Flagships Graphene Flagship & Human Brain Project - ERA-NETS 2019b (PeroGaS: MIS 5070514) is acknowledged for the financial support. The research work regarding the DFT calculations was supported by the Hellenic Foundation for Research and Innovation (HFRI) under the 3rd Call for HFRI PhD Fellowships (Fellowship Number: 5706). R.M. Giappa and I. Remediakis acknowledge computational time granted from the National Infrastructures for Research and Technology S.A. (GRNET S.A.) in the National HPC facility, ARIS, under the Projects NANOPTOCAT and COMPNANOMAT. We would like also to thank Mrs Alexandra Manousaki for the observation of the samples on SEM, Dr Emmanuel Spanakis for conducting the XPS measurements, Mr Lampros Papoutsakis and Aggeliki Sfakianou for their assistance to the temperature dependent XRD and gas sensing experiments in humidity, the Electron Microscopy Laboratory of the University of Crete for providing access to HRTEM and SEM facilities and the Material Science and Technology Department of the University of Crete for providing access to XPS facilities.

## Conflict of Interest

The authors declare no conflict of interest.

## Table of content

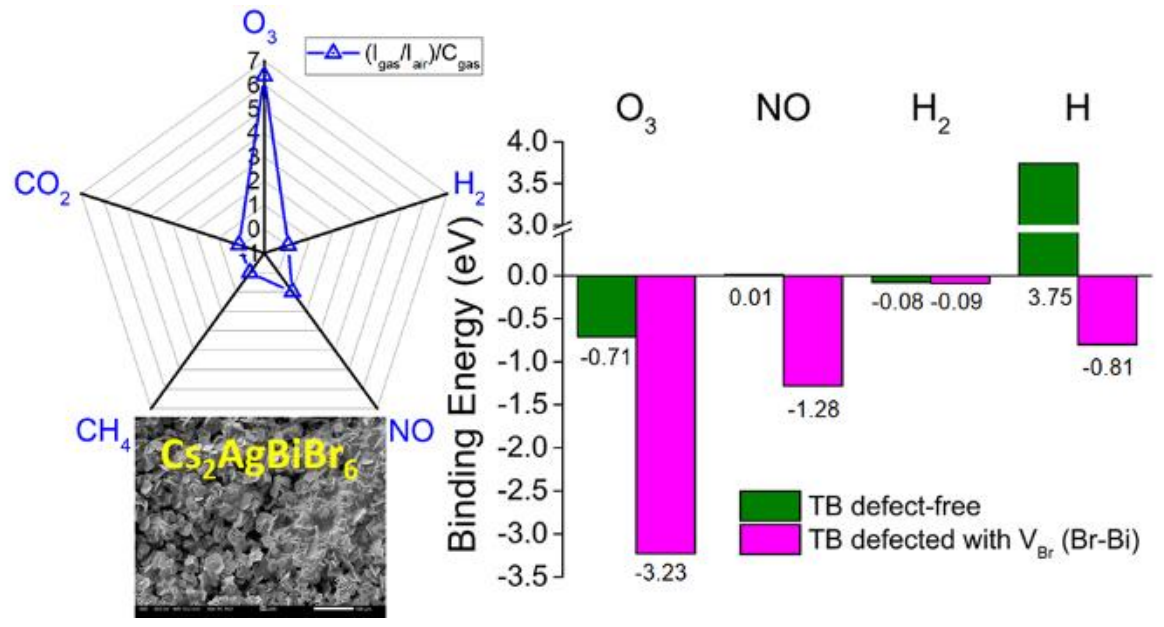

# Supporting Information

# Highly-stable, eco-friendly and selective $Cs_2AgBiBr_6$ perovskite-based ozone sensor

*Aikaterini Argyrou, Rafaela Maria Giappa, Emmanouil Gagaoudakis, Vassilios Binas, Ioannis Remediakis, Konstantinos Brintakis*, Athanasia Kostopoulou*, Emmanuel Stratakis**

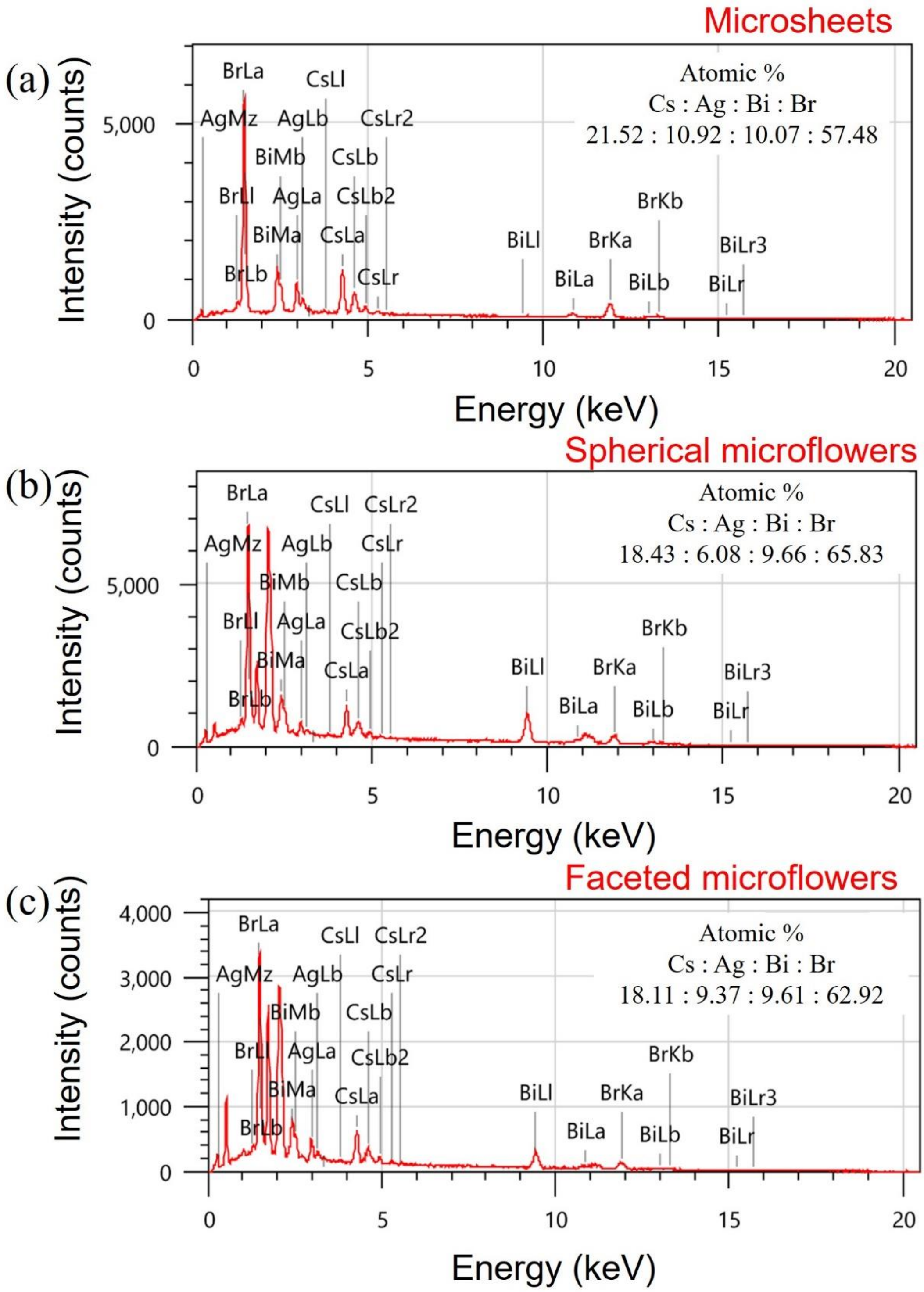

**Figure S1.** EDS analysis of $Cs_2AgBiBr_6$ microparticles of (a) sheet-, (b) faceted flower- and (c) spherical flower- shaped morphology.

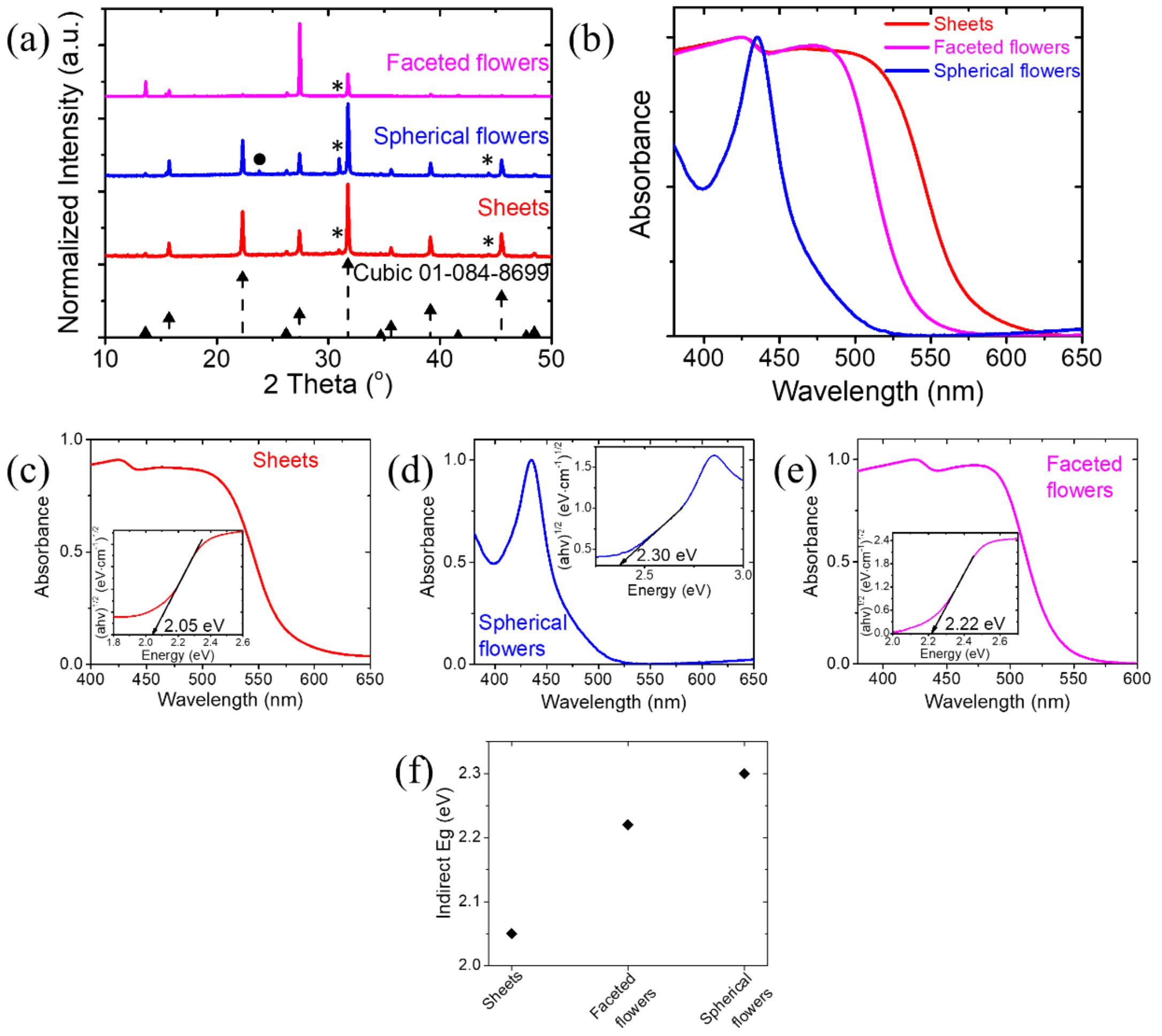

**Figure S2**. (a) XRD patterns of microsheet- (red pattern), spherical microflower- (blue pattern) and faceted microflower- (magenta pattern) based sensors. The asterisks (*) indicate the reflections of the side phases AgBr (COD 9008596) and the circles (●) show the $Cs_3Bi_2Br_6$ crystal structure (ICDD 01-070-0493). (b-e) Absorbance spectra of all the microcrystals and (f) the indirect band gap energy calculated through Tauc plots for indirect bandgap semiconductors of the same samples.

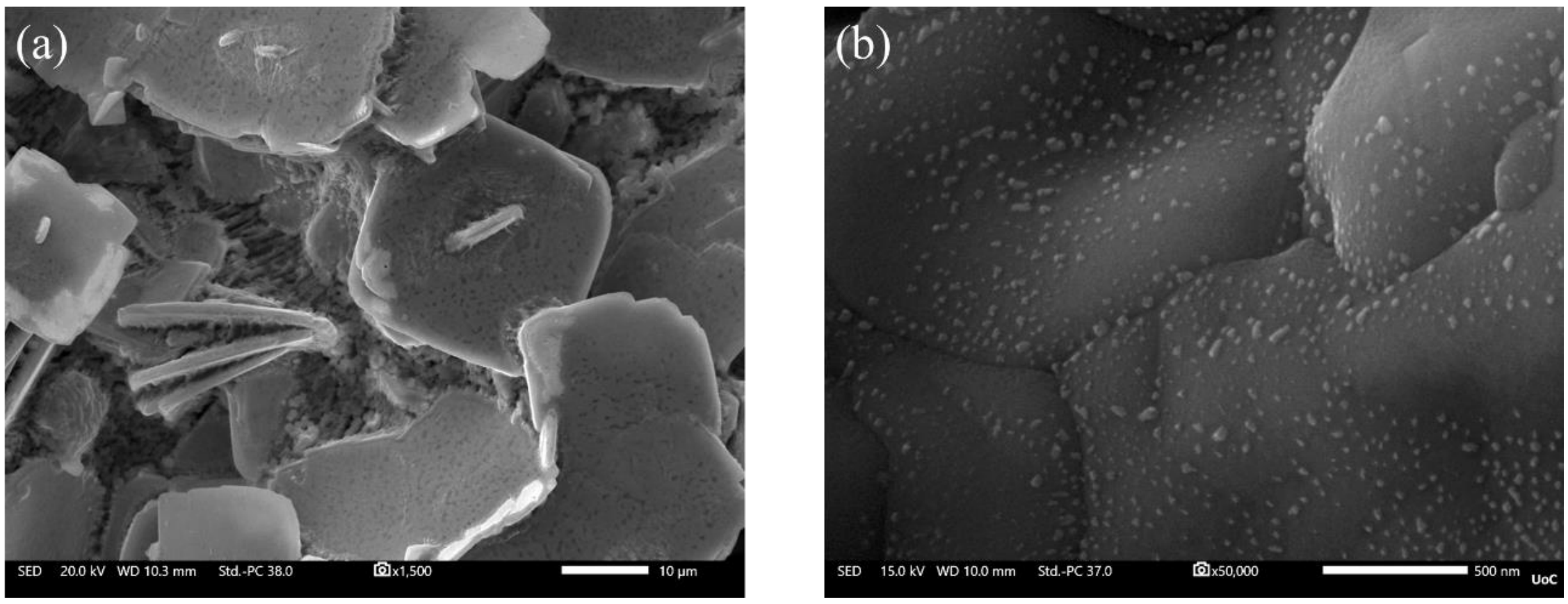

**Figure S3.** FESEM images of $Cs_2AgBiBr_6$ micro-sheets.

**Table S1.** $O_3$ sensing performance of different chemiresistive sensors.

| Sensing material | Concentration (ppb) | Response[a] | Ref. |
|---|---|---|---|
| $Cs_2AgBiBr_6$ microsheets | 168 | 1.07 | This work |
| ZnO | 30 | 44 (UV) | 1 |
| ZnO-gold | 30 | 108 (UV) | 1 |
| InGaZnO | 1000 | 13 (UV) | 2 |
| $SnO_2$-ZnO | 20 | 8 (UV) | 3 |
| $6SnO_2$ nanorods | 500 | 3 | 4 |
| 1wt% $AgIn_2O_3$ | 100 | 1.58 | 5 |
| 0.1% Au/$TiO_2$-$WO_3$ (3:1) | 2500 | 64 (LED) | 6 |
| NiAl | 15 | 1.22 | 7 |
| $In_2O_3$ | 10 | 4 (UV) | 8 |
| Au@$TiO_2$ | 500 | 1.15 (RH% 50) | 9 |
| $TiO_2$-$In_2O_3$ | 2000 | 56 (LED) | 10 |
| Carbon nanotubes | 5000 | 1.3 | 11 |

[a] Response = $R_{gas}/R_{air}$ or $R_{air}/R_{gas}$

**Table S2.** Metal halide perovskite ozone sensors at room temperature.

| Sensing material | Concentration (ppb) | Response [b] | $t_{res}$ (s) | $t_{rec}$ (s) | Ref. |
|---|---|---|---|---|---|
| $Cs_2AgBiBr_6$ microsheets | 168 | 1.07 | 37.2 | 82.8 | This work |
| $CsPbBr_3$ RC | 4 | 1.68 | 74.4 | 15.6 | [12] |
| $CsPbBr_3$ cubes | 187 | 1.56 | 143 | 320 | [13] |
| $CH_3NH_3PbI_{3-x}Cl_x$ thin film | 180 | 3 | 225 | 200 | [14] |

[b] Response $=\frac{I_{gas}}{I_{air}}$.

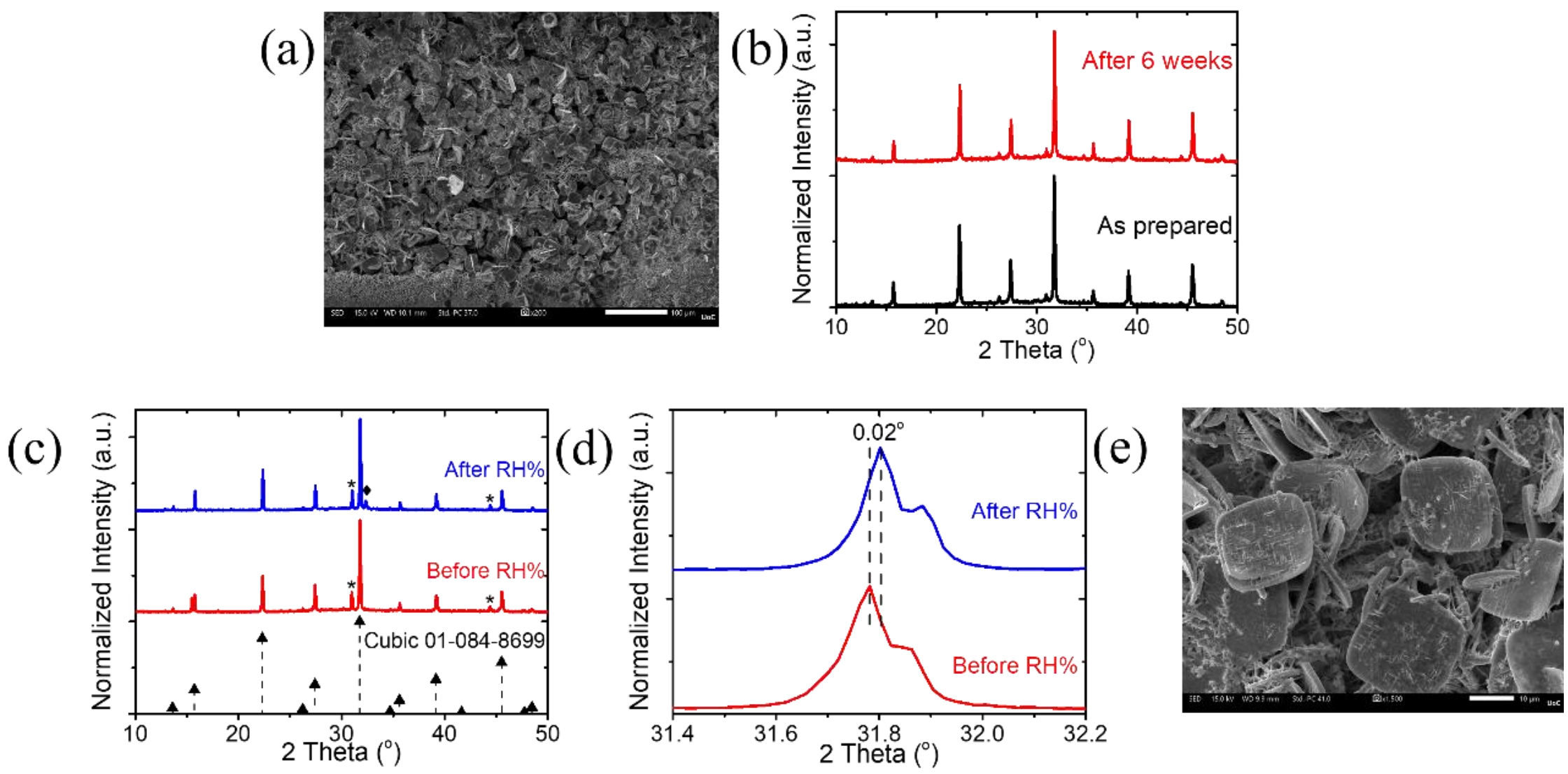

**Figure S4.** (a) FESEM image of $Cs_2AgBiBr_6$ microsheets at the end of the sensing process. (b) XRD patterns of $Cs_2AgBiBr_6$ microsheet-based sensor as prepared (black pattern) and after 6 weeks (red pattern). (c-d) XRD patterns as prepared of the microsheet-based sensors as prepared (red pattern) and after its exposure to $O_3$ gas under relative humidity (blue pattern). The asterisks (*) indicate the reflections of the side phases $CsAgBr_2$ (PDF #38-0850), AgBr (ICDD 00-006-0438) and Cs3Bi2Br6 (ICDD 01-070-0493) and the rhombus (♦) represents the formation of BiOBr (JCPDS 09-0393) (e) FESEM image of the microsheets after their exposure to RH%.

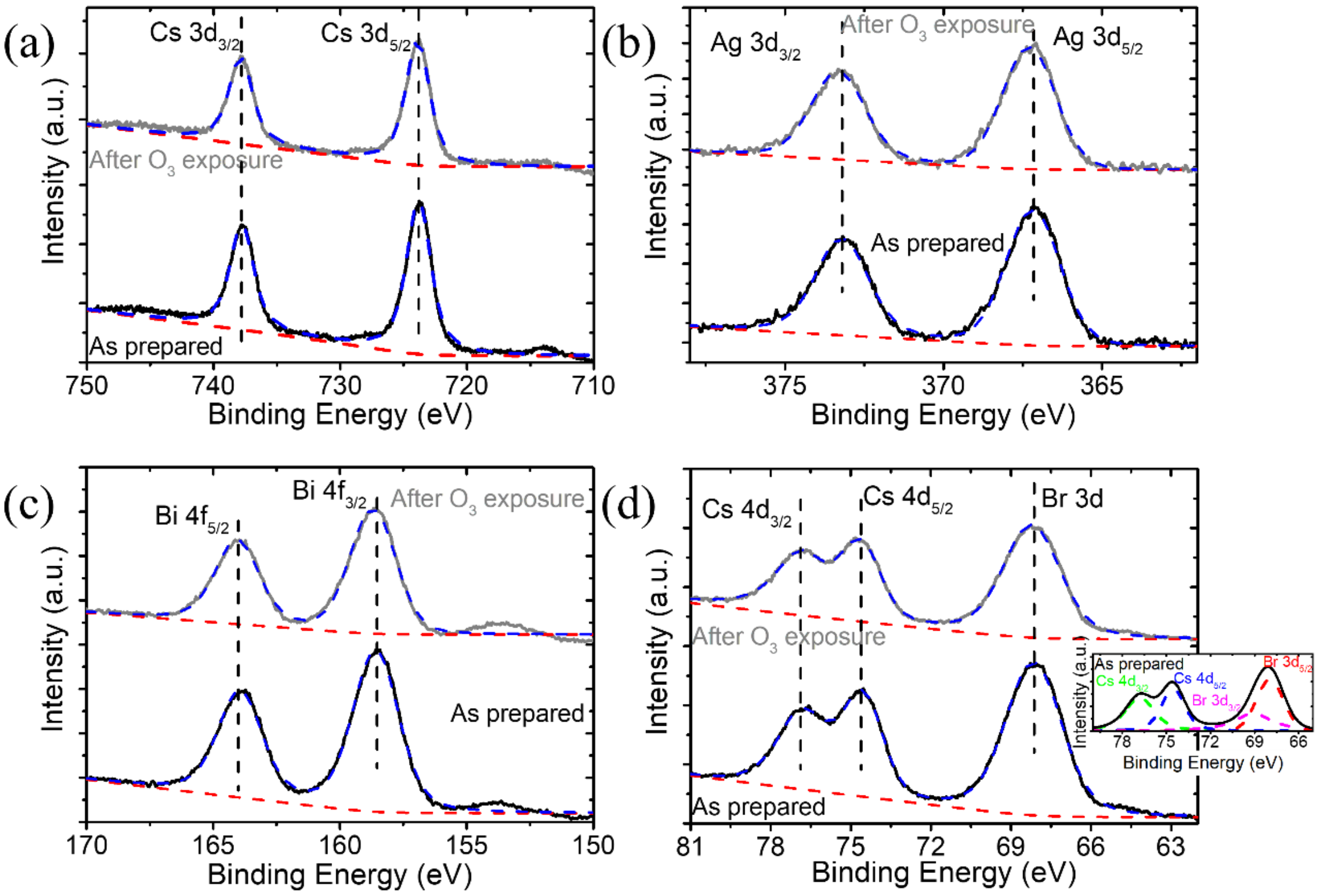

**Figure S5.** XPS narrow scans of (a) Cs 3d, (b) Ag 3d, (c) Bi 4f and (d) Br 3d of the $Cs_2AgBiBr_6$ micro-sheets before and after exposure to $O_3$. The inset shows the deconvolution of Br 3d orbitals of the as prepared sample with spin orbit splitting of 1.1 eV.

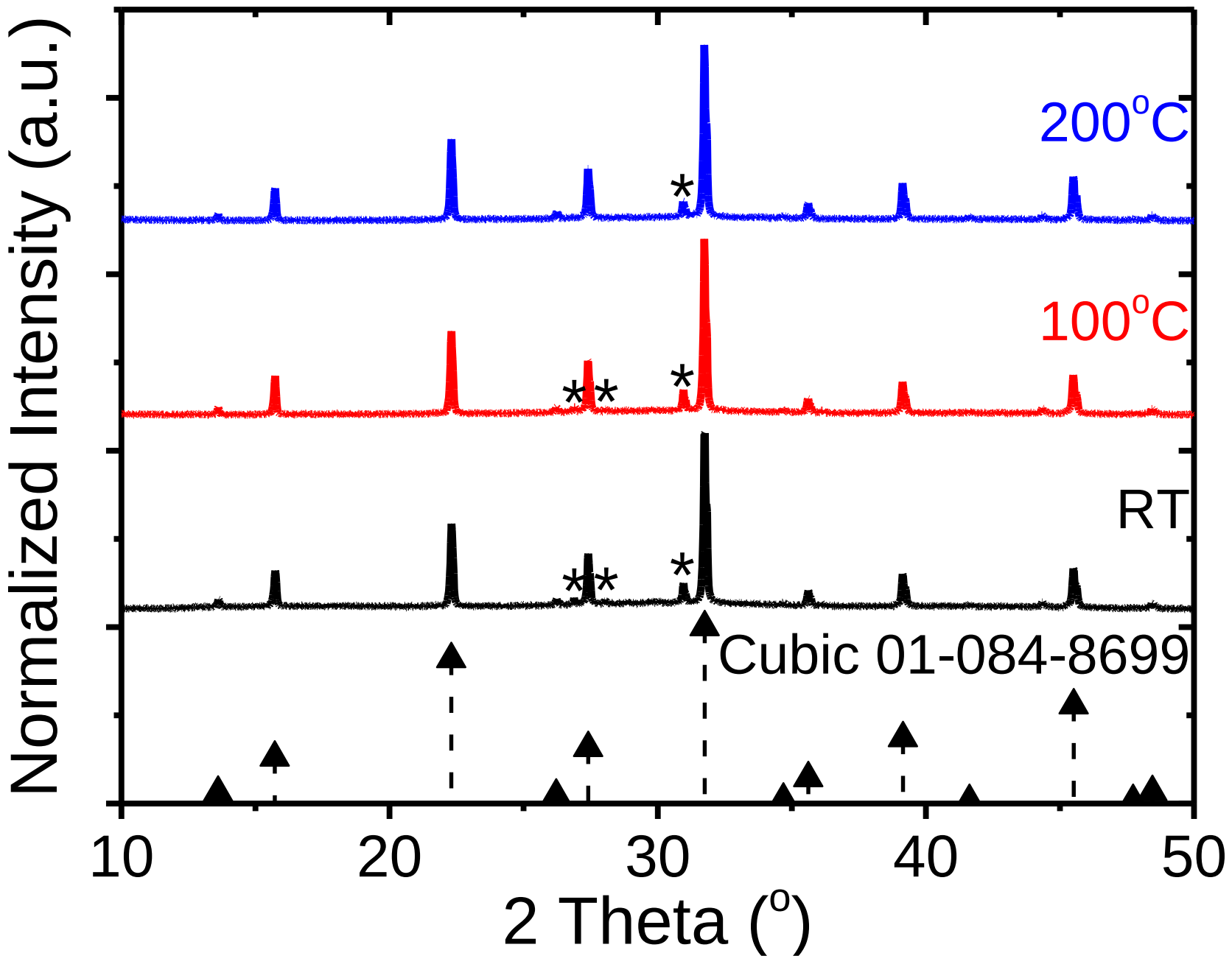

**Figure S6.** XRD patterns of the microsheets at the end of each annealing/$O_3$ sensing process. The asterisks correspond to the reflections of the side phases.

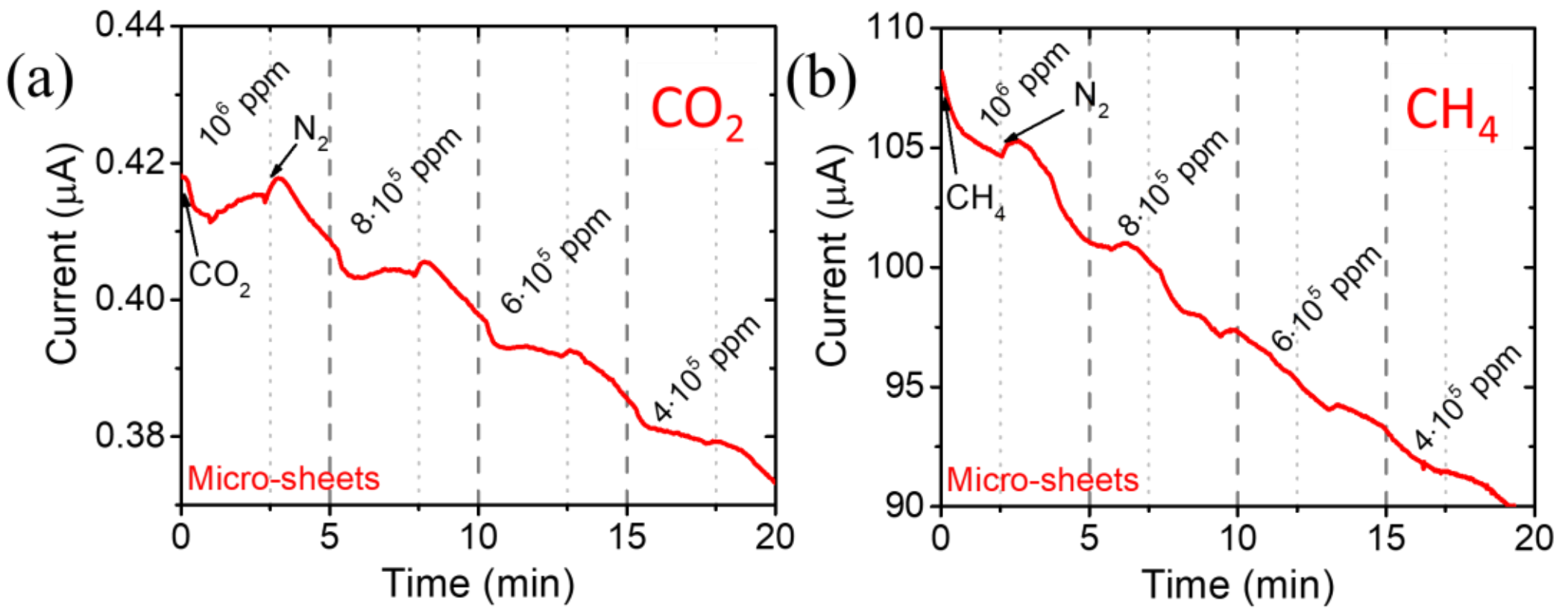

**Figure S7.** Electrical response curves of the microsheets-based sensor for (a) $CO_2$ and (b) $CH_4$ detection.

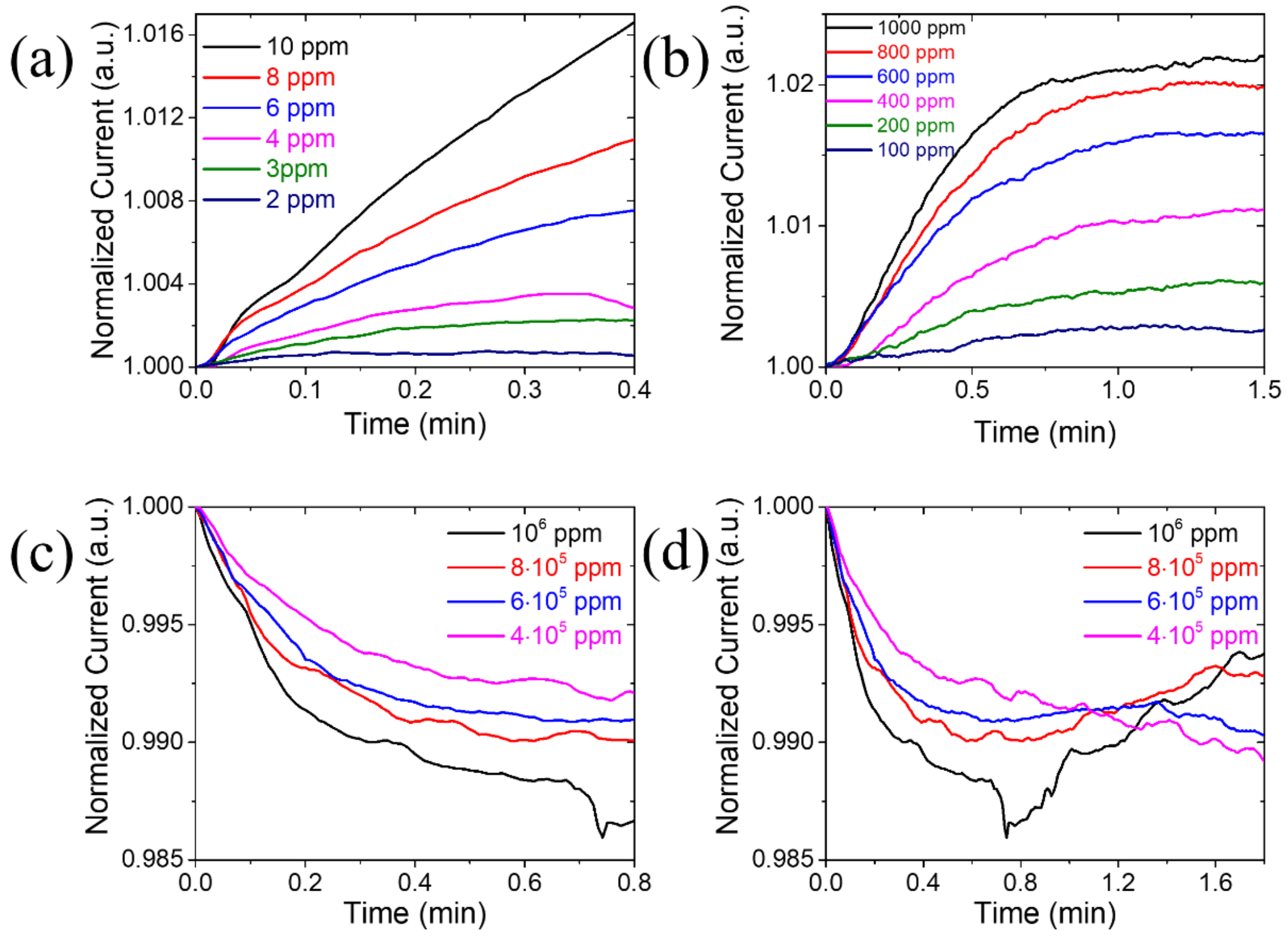

**Figure S8.** Normalized current curves upon (a) NO, (b) $H_2$, (c) $CH_4$ and $CO_2$ gas exposure of the $Cs_2AgBiBr_6$ microsheets as a function of time.

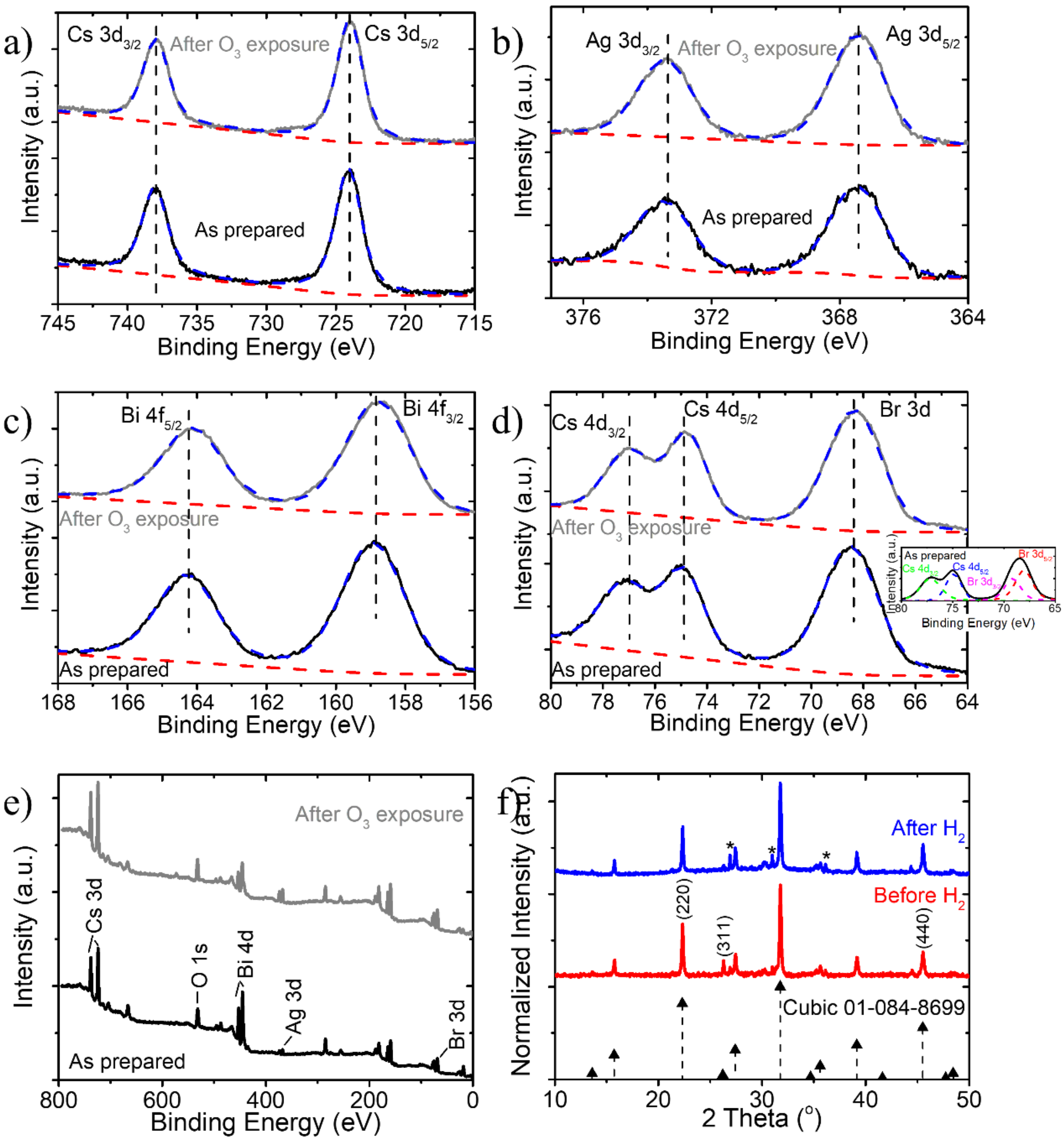

**Figure S9.** XPS narrow scans of (a) Cs 3d, (b) Ag 3d, (c) Bi 4f and (d) Br 3d of the$Cs_2AgBiBr_6$ microsheets before and after exposure to gas. The inset shows the deconvolution of Br 3d orbitals of the as prepared sample with spin orbit splitting of 1.1 eV. (e) Survey spectra of $H_2$ sensors as prepared (lower, black curve) and at the end of the hydrogen process (upper, grey curve) (f) XRD patterns of $Cs_2AgBiBr_6$ microsheet-based sensor as prepared (red) and after its exposure to $H_2$. The asterisks (*) indicate the reflections of side phases.

**DFT calculations**

## I. Adsorption Energies

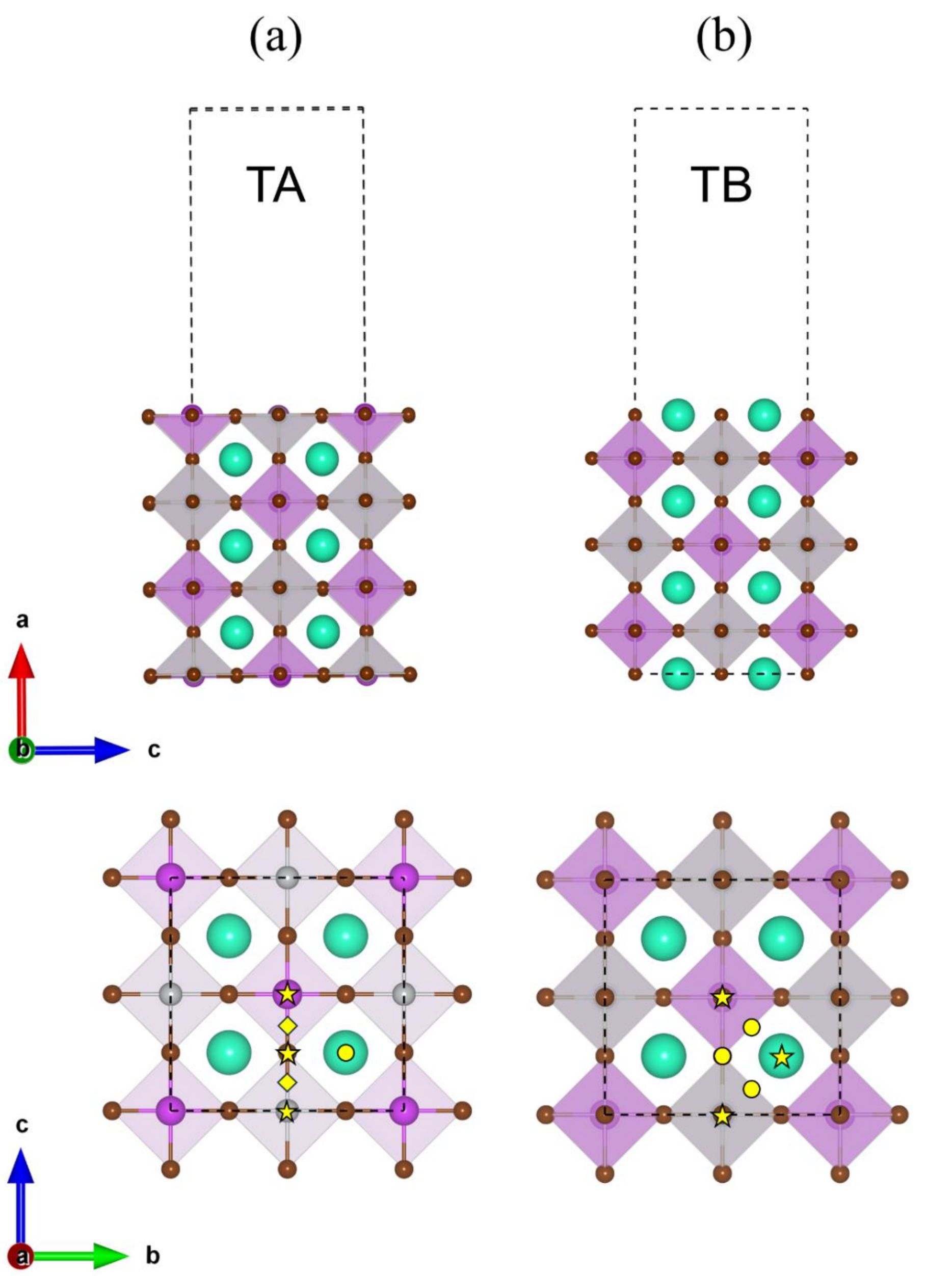

**Figure S10.** The 7-layer slab model with (a) the BiBr/$AgBr_3$ (TA) termination, and (b) the CsBr termination (TB), viewed along the b-axis in the top and the a-axis in the bottom panels, respectively. The different adsorption site types, i.e., on-top, bridge, hollow, are denoted with yellow stars, diamonds, and circles, respectively.

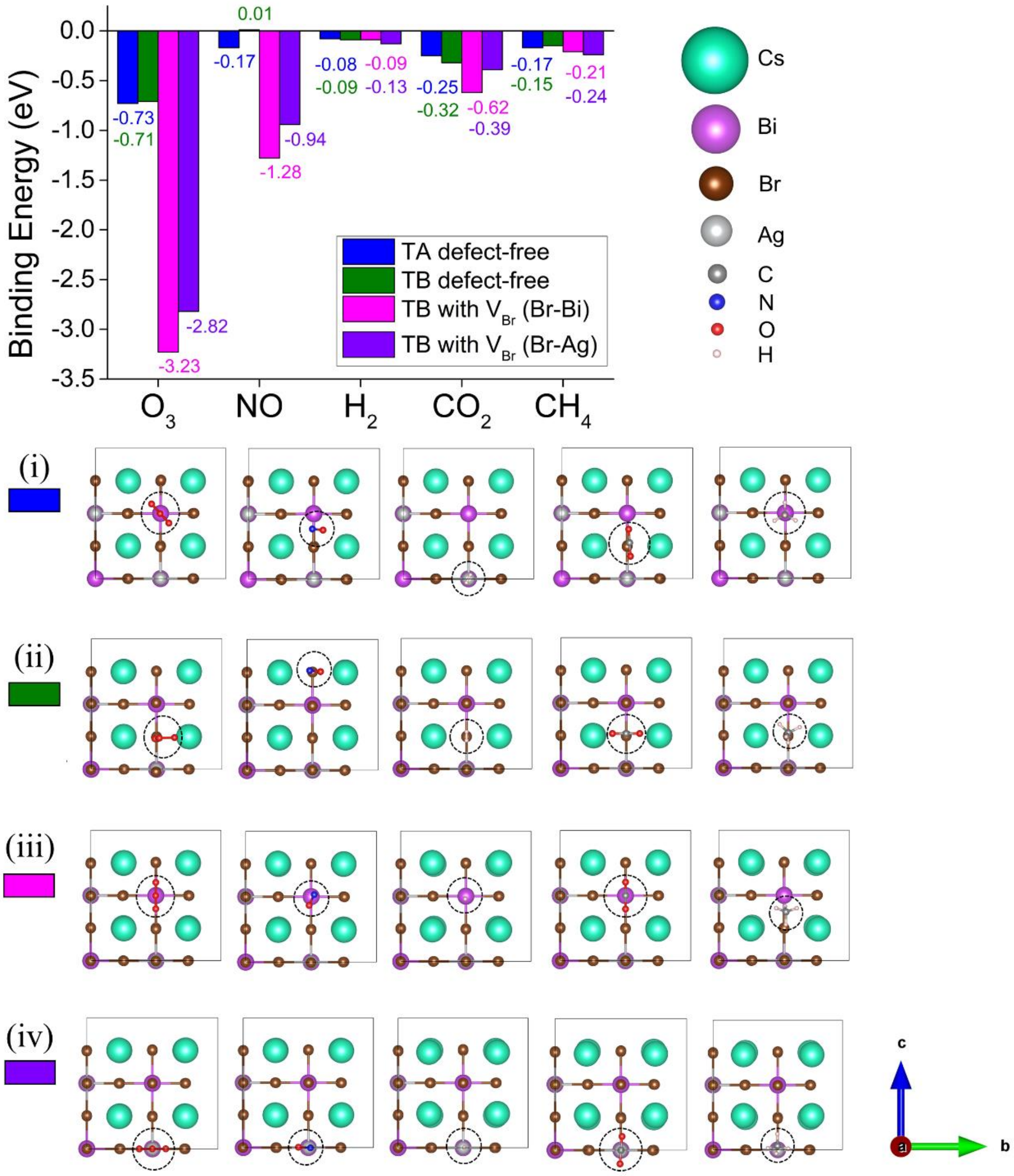

**Figure S11.** Adsorption energy in the strongest binding site of each target molecule on the defect-free TA (blue) and TB (green) surface, on the defected TB with $V_{Br}$ (Br-Bi) (magenta) and with $V_{Br}$ (Br-Ag) (violet) surface. The strongest binding sites of each target molecule (i) on the TA surface without defects, (ii) on the TB surface without defects, (iii) on the defected TB surface with $V_{Br}$ (Br-Bi), and on the (iv) defected TB surface with $V_{Br}$ (Br-Ag). Cs, Ag, Bi, Br, O, N, C, and H atoms are illustrated as cyan, silver, purple, brown, red, blue, grey, and white spheres, respectively.

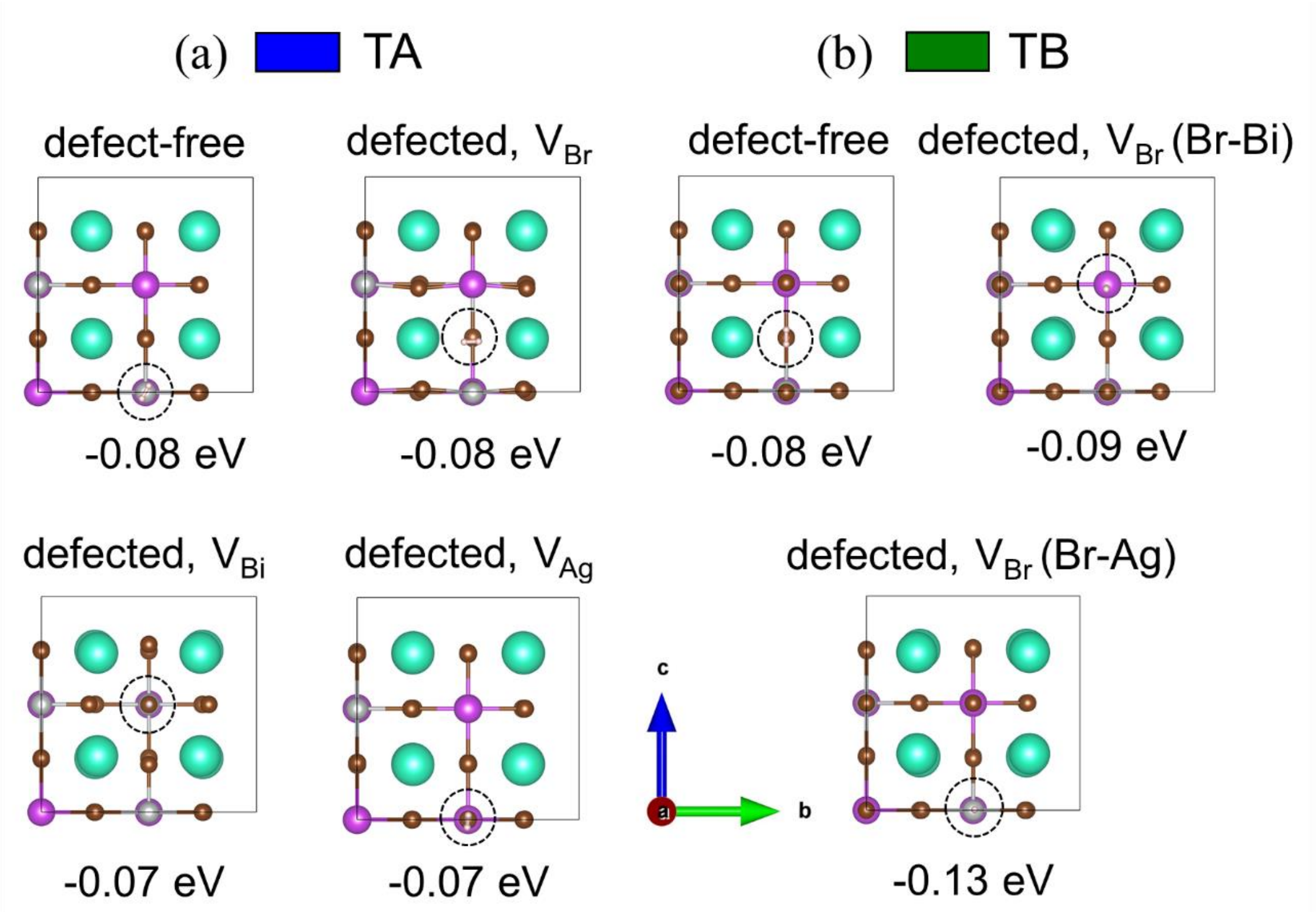

**Figure S12.** The energetically favorable adsorption sites for the hydrogen molecule on the (a) TA defect-free and defected with $V_{Br}$, $V_{Bi}$, $V_{Ag}$ surface, and on the (b) TB defect-free and defected with $V_{Br}$ (Br-Bi) and $V_{Br}$ (Br-Ag), along with their corresponding binding energies in eV. Cs, Ag, Bi, Br, and H atoms are illustrated as cyan, silver, purple, brown, and white spheres, respectively.

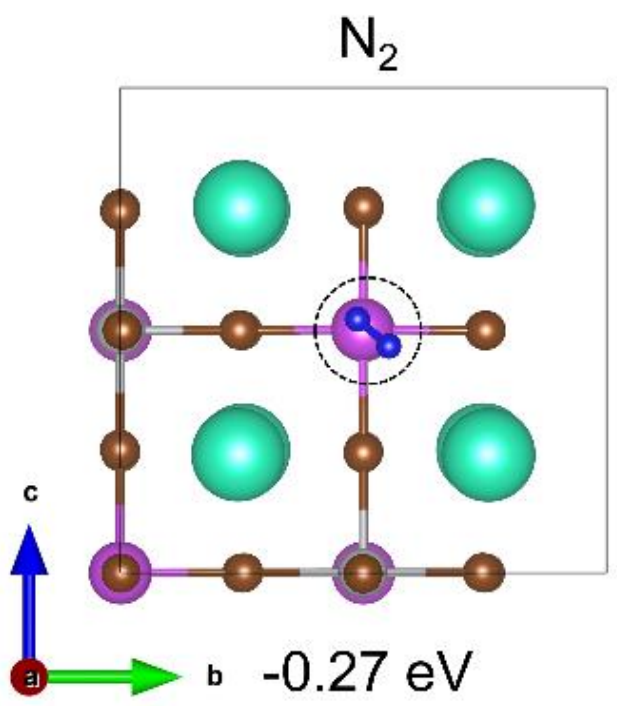

**Figure S13.** The energetically favorable adsorption site for $N_2$ on the defected TB surface with $V_{Br}$ (Br-Bi), along with the corresponding binding energy in eV. Cs, Ag, Bi, Br, and N atoms are illustrated as cyan, silver, purple, brown, and blue spheres, respectively.

**II. Density of States**

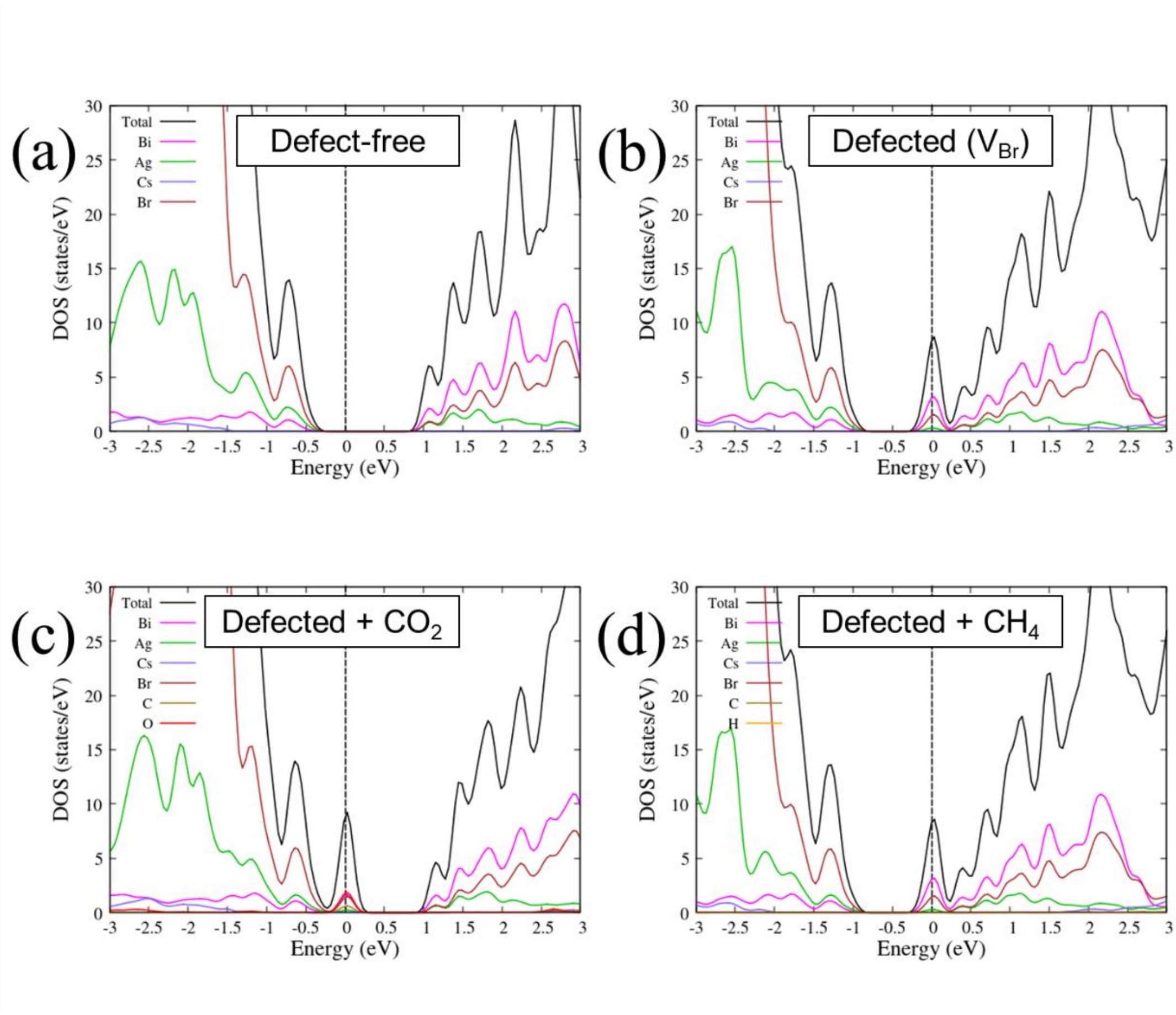

**Figure S14.** Partial density of states (PDOS) of the (100) oriented TB $Cs_2AgBiBr_6$ model surfaces (a) without defects and (b) defected with $V_{Br}$ (Br-Bi). PDOS of the defected with $V_{Br}$ (Br-Bi) TB surface upon adsorption of (c) $CO_2$, and (d) $CH_4$.

Upon adsorption of the $CO_2$ molecules in the vacancy site, the defected states vanish and new distinct states appear around the Fermi level, with contributions from O, Bi and Br atoms being nearly equivalent and C atoms contributing slightly less. On the other hand, the defected states persist in the case of the weakly interacting $CH_4$.

## III. Charge Density Difference Plots

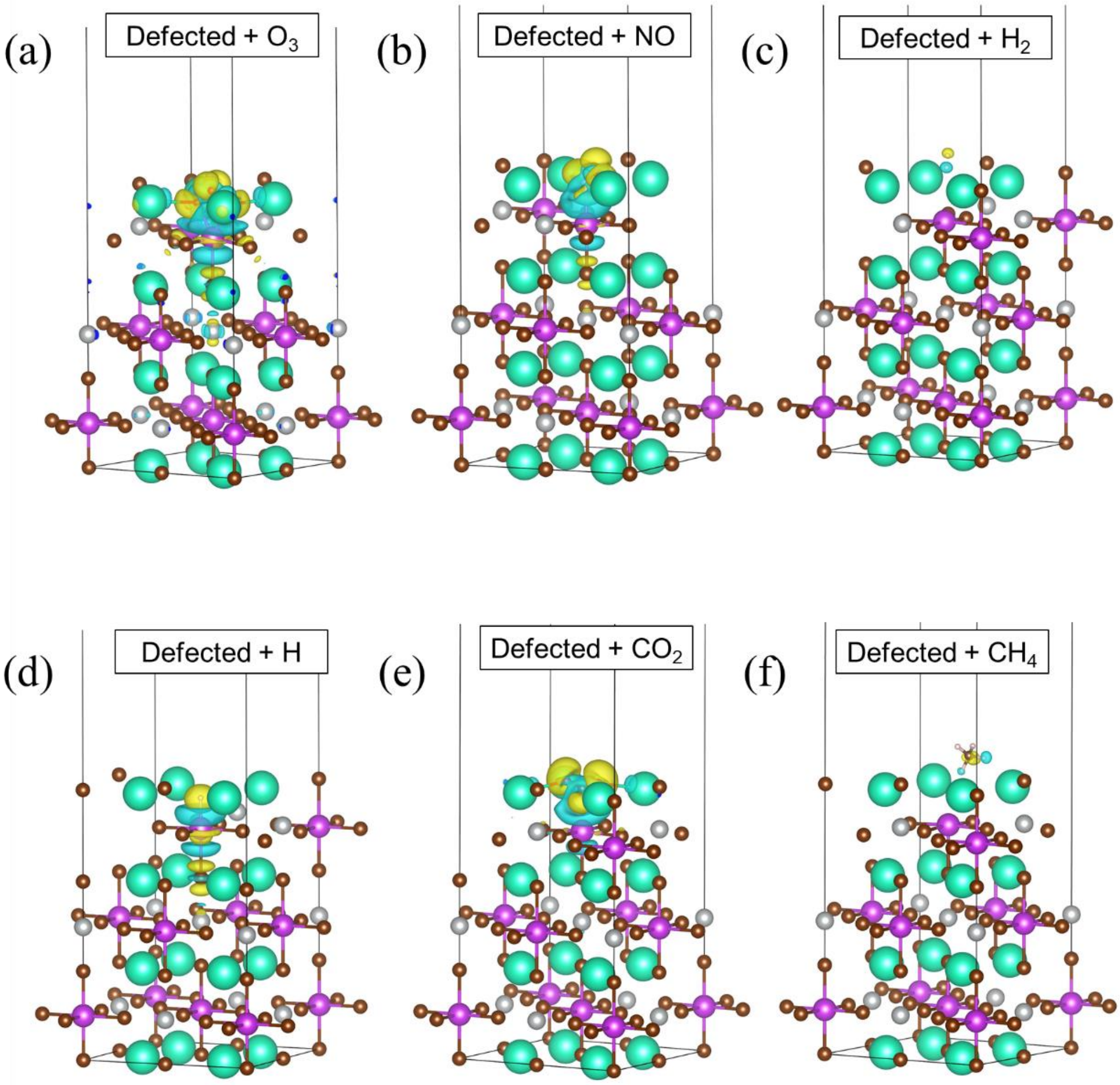

**Figure S15.** Charge density difference plots calculated at 0.001 (e/bohr$^3$) isosurface for (a) $O_3$, (b) NO, (c) $H_2$, (d) H, (e) $CO_2$, and (f) $CH_4$ adsorbed on the defected TB surface with $V_{Br}$ (Br-Bi). Yellow and cyan areas illustrate charge accumulation and depletion, respectively. Cs, Ag, Bi, Br, O, C, N, and H atoms are illustrated as cyan, silver, purple, brown, red, blue, grey and white spheres, respectively.

In all strongly interacting cases, there is pronounced polarization effect in the upper layers of the surface around the adsorption site, indicative of the rearrangement of the electron density in response to the presence of the adsorbate molecules. At the same time, there is significant polarization of the electron density associated with the adsorbate atoms themselves, suggesting significant charge transfer from the uppermost surface atoms to the adsorbed molecules. On the other hand, for the weakly interacting $H_2$ and $CH_4$ cases, minimal to no polarization effects are observed in the surface and the charge density distribution of the adsorbates remains largely unperturbed.